\documentclass[letter]{article}

\usepackage{amsthm}
\usepackage{times}
\usepackage{url}
\usepackage{epsfig}
\usepackage{amsmath}
\usepackage{longtable}
\usepackage{amsfonts}
\usepackage{amssymb}
\usepackage{graphicx}
\usepackage{epstopdf}
\usepackage{multirow}
\usepackage{subcaption}
\usepackage{verbatim}
\usepackage{enumitem}
\usepackage{color}
\usepackage{hyperref}
\usepackage{xspace}
\usepackage{gensymb}
\usepackage{setspace}
\usepackage{booktabs}

\usepackage{float}
\restylefloat{figure}

\usepackage[linesnumbered,ruled]{algorithm2e}

\usepackage{self}

\newtheorem{example}{Example}

\usepackage{hyperref}
\usepackage{xcolor}
\hypersetup{
	colorlinks,
	linkcolor={red!50!black},
	citecolor={blue!50!black},
	urlcolor={blue!80!black}
}

\usepackage[numbers]{natbib}

\graphicspath{{img/}}

\begin{document}

\title{High-Resolution Traffic Sensing with Autonomous Vehicles}

\author{Wei Ma, Zhen (Sean) Qian\\
	Department of Civil and Environmental Engineering\\ Carnegie Mellon University, Pittsburgh, PA 15213\\
	\{weima, seanqian\}@cmu.edu}
\maketitle

\begin{abstract}
The last decades have witnessed the breakthrough of autonomous vehicles (AVs), and the perception capabilities of AVs have been dramatically improved. Various sensors installed on AVs, including, but are not limited to, LiDAR, radar, camera and stereovision, will be collecting massive data and perceiving the surrounding traffic states continuously. In fact, a fleet of AVs can serve as floating (or probe) sensors, which can be utilized to infer traffic information while cruising around the roadway networks. In contrast, conventional traffic sensing methods rely on fixed traffic sensors such as loop detectors, cameras and microwave vehicle detectors. Due to the high cost of conventional traffic sensors, traffic state data are usually obtained in a low-frequency and sparse manner. In view of this, this paper leverages rich data collected through AVs to propose the high-resolution traffic sensing framework. The proposed framework estimates the fundamental traffic state variables, namely, flow, density and speed in high spatio-temporal resolution, and it is developed under different levels of AV perception capabilities and low AV market penetration rate. The Next Generation Simulation (NGSIM) data is adopted to examine the accuracy and robustness of the proposed framework. Experimental results show that the proposed estimation framework achieves high accuracy even with low AV market penetration rate. Sensitivity analysis regarding AV penetration rate, sensor configuration, and perception accuracy will also be studied. This study will help policymakers and private sectors (e.g Uber, Waymo) to understand the values of AVs, especially the values of massive data collected by AVs, in traffic operation and management.
\end{abstract}
\section{Introduction}

As the combination of a wide spectrum of cutting-edge technologies, autonomous vehicles (AVs) are destined to fundamentally change and reform the whole mobility system \citep{litman2017autonomous}. AVs have great potentials in improving safety and mobility \citep{assidiq2008real, tientrakool2011highway, stern2018dissipation}, reducing fuel consumption and emission \citep{vahidi2018energy, gawron2018life}, and redefining the civil infrastructure systems such as road networks \citep{chen2016optimal, chen2017optimal, duarte2018impact}, parking spaces \citep{zhang2015exploring, harper2018exploring, millard2019autonomous}, and public transit systems \citep{lutin2018not, salonen2018passenger}. 
Over the past two decades, many advanced driver assistance systems (ADAS) ({\em e.g.} lane keeping, adaptive cruise control) have been deployed in various type of production vehicles. Currently, both traditional car manufacturers and high-tech companies are competing to lead the full autonomy (L4$\sim$L5) technologies. For example, Waymo’s AVs alone are driving 25,000 miles every day in 2018 \citep{waymo2} and there have been commercialized AVs operating in multiple cities by Uber \citep{uberatg}.

Despite of the rapid development of AVs technologies, there is still a long way to reach the full autonomy and to completely replace all conventional vehicles with AVs.  We will witness a long period over which AVs and conventional vehicles co-exist on public roads. How to sense, model and manage the mixed transportation systems presents a great challenge to public agencies. To the best of our knowledge, most current studies view AVs as controllers and focus on modeling and managing the mixed traffic networks \citep{zhao2019enhanced}. For example, novel System Optimal (SO) and User Equilibrium (UE) models are established to include AVs \citep{levin2017congestion, wang2019multiclass}, coordinated intersections are proposed to improve the traffic throughput \citep{shida2009development,li2018piecewise, yu2018integrated}, vehicle platooning strategies are developed to reduce highway congestion \citep{li2015overview, gong2016constrained}, and AVs can also complement conventional vehicles to solve last-mile problems \citep{chong2011autonomous, moorthy2017shared}. However, there is a lack of studies in traffic sensing methods for the mixed traffic networks.

In this paper, we advocate the great potentials of AVs as moving observers in high-resolution traffic sensing. We note that traffic sensing with AVs in this paper is different from perception of AVs \citep{van2018autonomous}. The perception of AV is the key to the safe and reliable AVs, and it refers to the ability of AVs to collect information and extract relevant knowledge from the environment using various sensors \citep{pendleton2017perception}. While traffic sensing with AVs refers to estimating the traffic conditions, such as flow, density and speed using the information perceived by AVs \citep{seo2017traffic}. To be precise, traffic sensing with AVs is built on top of the perception technologies on AV, and in this paper we will discuss the impact of different perception technologies on traffic sensing.

In fact, a fleet of autonomous vehicles (AVs) can serve as floating (or probe) sensors, detecting and tracking the surrounding traffic conditions to infer traffic information while cruising around the roadway network. Enabling traffic sensing with AVs is cost effective. The AVs equipped with various sensors and data analytics capabilities may be costly. While costly, those sensors and data are used primarily to detect and track adjacent objects to enable safe AV driving in the first place. Therefore, there is no additional overhead cost of these data collections for traffic sensing since it is a secondary use. 

High-resolution traffic sensing is central to traffic management and public policies. For instance, local municipalities would need information regarding how public space ({\em e.g.} curbs) is being utilized to set up optimal parking duration limits; Metropolitan planning agencies would need various types of traffic/passenger information, including travel speed, traffic density and traffic flow by vehicle classifications, as well as pedestrians and cyclists. Infrastructure planning requires intensive spatial and temporal coverage, and having data only at sparse locations on highways is far from being sufficient; Those data would also be essential to design optimal traffic signal timing and schedule infrastructure maintenance. In addition, non-emergent and emergent incidents are reported by citizens through 911 system, respectively. Automated traffic sensing, both historical and in real time, can complement those systems to enhance their timeliness, accuracy and accessibility. In general, accurate and ubiquitous information of infrastructure and usage patterns in  public space are currently missing.

By leveraging the rich data collected through AVs, we are able to detect and track various objects in transportation networks. The objects include, but are not limited to, moving vehicles by vehicle classifications, parked vehicles, pedestrians, cyclists, signage in public space. When all those objects in high spatio-temporal resolutions are being continuously tracked, those data can be translated to useful traffic information for public policies and decision making. The three key features of traffic sensing based on autonomous vehicles sensors are: inexpensive, ubiquitous and reliable. Those data are collected by automotive manufacturers for guiding autonomous driving in the first place, which promises great scalability in this approach. With minimum additional efforts, the same data can be effectively translated into information useful for the community. For instance, how much time in public space at a particular location is utilized by different classifications of vehicles and by what travel modes, respectively? Can we effectively evaluate the accessibility, mobility and safety of the mobility networks? The sensing coverage will become ubiquitous in the near future, provided with an increasing market share of autonomous vehicles. Data acquired from individual autonomous vehicles can be compared, validated, corrected, consolidated, generalized and anonymized to retrieve most reliable and ubiquitous traffic information. In addition, this paper for traffic sensing also implies the future possibility of interventions for effective and timely traffic management. It enables real-time traffic monitoring, potentially safer traffic operation, faster emergency response, and smarter infrastructure management. 

The rest of this paper focuses on a critical problem to estimate the fundamental traffic state variables, namely, flow, density and speed, in high resolution to demonstrate the sensing power of AVs. In addition to traffic sensing, there are many aspects and data in community sensing that could be explored in the near future. For example, perception of AVs can be used for monitoring urban forest health, air quality, street surface roughness and many other applications of municipal asset management \citep{ma2019measuring, xu2019ilocus, mahmoudzadeh2019estimating}.

Traffic state variables ({\em e.g.} flow, density and speed) play a key role in traffic operation and management. Over the past several decades, traffic state estimation (TSE) methods have been developed for not only stationary sensors ({\em i.e.} Eulerian data) but also moving observers ({\em i.e.} Lagrangian data) \citep{sun2017simultaneous}. Stationary sensors, including loop detectors, cameras and radar, monitor the traffic conditions at a fixed location. Due to the high installation and maintenance cost, the stationary sensors are usually sparsely installed in the network, and hence the collected data are not sufficient for the practical traffic operation and management \citep{jain2019review}. Data collected by moving observers ({\em e.g.} probe vehicles, ride-sourcing vehicles, unmanned aerial vehicles, mobile phones) has a better spatial coverage and hence it enables cost-effective TSE in large-scale networks \citep{antoniou2011synthesis}. 
Though the TSE method with moving observer can date back to 1954 \citep{wardrop1954method}, recent advances in communication and Internet of Things (IoT) technologies have catalyzed  the development and deployment of various moving observers in real-world. Readers are referred to \citet{wang2005real, seo2017traffic} for a comprehensive review of existing TSE models. 

To highlight our contributions, we present studies that are closely related to this paper. The moving observers can be categorized into four types: originally defined moving observers, probe vehicles (PVs), unmanned aerial vehicles (UAVs) and AVs. Their characteristics and related TSE models are presented as follows: 

\begin{itemize}
	\item {\em Originally defined moving observers.} The moving observer method for TSE is originally proposed by \citet{wardrop1954method}. The proposed method requires a probe vehicle to transverse along the road and count the number of slower vehicles overtaken by the probe vehicle and the number of faster vehicles which overtake the probe vehicle \citep{wright1973theoretical}. Though the setting of the originally define moving observers is too ideal for practice, it enlightened us on the value of using Lagrangian data for TSE.
	\item {\em PVs.} The PVs refer to all the vehicles that can be geo-tracked, and it includes, but is not limited to, taxis, buses, trucks, connected vehicles, ride-sourcing vehicles \citep{zheng2015trajectory}. The PV data has great advantages in estimating speed, while it hardly contains density/flow information. Studies have explored the sensing power of PVs \citep{o2019quantifying}. PV data is usually used to complement stationary sensor data to enhance the traffic state estimation \citep{herrera2010incorporation, van2018macroscopic}. PVs with spacing measurement equipment can estimate traffic flow and speed simultaneously \citep{wilby2014lightweight, seo2015traffic, seo2015estimation, fountoulakis2017highway}. 
	\item {\em UAVs.}  By flying over the roads and viewing from top-view perspectives, UAVs are able to monitor a segment of road or even the entire network \citep{puri2005survey,kanistras2015survey, ke2018real}. UAVs have the advantage of better spatial coverage while extra purchase of UAVs and the corresponding maintenance cost are required. Traffic sensing with UAV has been extensively studied in recent years, including vehicle identification algorithms \citep{zhu2018urban, khan2018unmanned, ke2018real}, sensing frameworks \citep{jin2016unmanned, niu2018uav}, and UAV routing mechanisms \citep{li2018unmanned, liu2019real}.
	\item {\em AVs.} AVs can be viewed as probe vehicles equipped with more sensors and hence have better perception capabilities. Not only the AV itself can be geo-tracked, the vehicles surrounded by AVs can also be detected and tracked. AVs also share some similarities with UAVs because AVs can scan a continuous segment of road. We believe AVs fall in between the PVs and UAVs, and hence existing TSE methods can hardly be applied to AVs. Furthermore, there are few studies on TSE with AVs. \citet{chen2017cyber} presents a cyber-physical system to model the traffic flow near AVs based on flow theory, while the TSE for the whole road is not studied. Recently, Uber ATG conduct an experiment to explore the possibility of TSE using AVs \citep{tseav}. 
\end{itemize} 

Given the unique characteristics of AVs as moving observers, there is a great need to study the AV-based TSE methods. In view of this, we develop a data-driven framework that estimates high-resolution traffic state variables, namely flow, density and speed using the massive data collected by AVs. The framework clearly defines the task of TSE with AVs involved and considers different perception levels of AVs. A two-step TSE method is proposed under a low AV market penetration rate. The main contributions of this paper are summarized as follows:

\begin{itemize}
	\item It discusses the functionality and role of various sensors in traffic state estimation. The sensing power of AVs is categorized into three levels. 
	\item It builds a two-step framework that leverages the sensing power of AVs to estimate high-resolution traffic state variables. The first step directly translates the information observed by AVs and the second step employs data-driven methods to estimate the information that is not observed by AVs. The proposed estimation methods are data-driven and can be interpreted by the traffic flow theory. 
	\item  The Next Generation Simulation (NGSIM) data is adopted to examine the accuracy and robustness of the proposed framework. Experimental results are compelling, satisfactory and interpretable. Sensitivity analysis regarding AV penetration rate, sensor configuration, and perception accuracy will also be studied.
\end{itemize}

The remainder of this paper is summarized as follows. Section~\ref{sec:sensing} discusses the sensing power of AVs. Section~\ref{sec:for} rigorously formulates the high-resolution TSE framework with AVs, followed by a discussion of the solution algorithms in section~\ref{sec:sol}. In section~\ref{sec:exp}, numerical experiments are conducted with NGSIM data to demonstrate the effectiveness of the proposed framework. Lastly, conclusions are drawn in section \ref{sec:con}.

\section{Sensing power of autonomous vehicles}
\label{sec:sensing}
In this section, we will discuss different levels of AV perception capabilities and how they associate with traffic sensing. We first discuss various sensors installed on AVs and their relation to traffic sensing. Analogous to the automation level definitions from Society of Automotive Engineers (SAE), we define three sensing levels of AVs. Lastly, we discuss a conceptual data center for processing the sensing data.

\subsection{Sensors}
In this section, we discuss different types of sensors used for AV perception and their potential usage for traffic sensing. Sensors for perception that are mounted on AVs include, but are not limited to, camera, stereo vision camera, LiDAR, radar and sonar \citep{pendleton2017perception}.

A camera can detect shapes and colors, so it is widely used for object detection ({\em e.g.} signals, pedestrians, vehicles and lane marks). Due to its low cost, multiple cameras can be mounted on a single AV. Theoretically, studies have shown that camera data can be used for object detection, tracking and traffic sensing \citep{shan2015camera, bautista2016convolutional}. In practice, camera image does not contain depth (distance) information, localization of vehicles is challenging for a single camera. On the modern AV prototypes, cameras are usually fused with stereo vision camera system or LiDAR to perceive the surrounding environments. In particular, the shape and color information obtained from camera is essential for object tracking \citep{aly2008real, dollar2009pedestrian}. Stereo vision camera refers to a device with two or more cameras horizontally mounted. Stereo vision camera is able to obtain the depth information of each pixel from the slightly different images taken by its cameras. 

Light Detection and Ranging (LiDAR) uses the pulsed laser beam to measure the distance between the detected object and itself. LiDAR can also obtain the 3D shape of the detected object. The LiDAR used on AVs is typically 360\degree, and the detection range varies from 30 to 150 meters, depending on makers, detection algorithms and weather conditions. Both LiDAR and stereo vision camera can be used for vehicle detection and 3D mapping. The system latency (time delay for processing the retrieved data) of stereo vision camera is higher than LiDAR, though the price of stereo vision camera is much cheaper \citep{van2018autonomous}. Theoretically either of the LiDAR or stereo vision camera can be used to build the full AV perception system, while currently most of AVs use LiDAR as the primary sensor.

There are two types of radar mounted on AVs. 
The short-range radar (SRR) is typically used for blind spot detection, parking assist and collision warning. The range for SRR is around $20$ meters \citep{takatori2006stand}.  Similarly, sonar, with its limited detection range (3 to 5 meters), is also frequently used for blind spot detection and parking assist.  Neither of the two sensors are considered as appropriate sensors for traffic sensing. In contrast, the long-range radar (LRR), which is primarily used for adaptive cruise control, can be potentially used for traffic sensing. The range for LRR is around $150$ meters and it is dedicated to detect the preceding vehicle in its current lane.

To conclude, Table~\ref{tab:sensor} summarizes a list of sensors that can be potentially used for traffic sensing based on above discussions and \citet{Thakur2018,van2018autonomous}.

\begin{table}[h]
	\caption{Sensors used for traffic sensing}
	\centering
	\label{tab:sensor}
	\begin{tabular}{|c|c|c|}
		\hline
		Sensors & Usage & Range  \\\hline\hline
		Camera & Surrounding vehicle detection/tracking, lane detection & $20\sim60$ meters \\
		Stereo vision camera & Surrounding vehicle detection/tracking, 3D mapping & $20\sim60$ meters \\
		LiDAR & Surrounding vehicle detection/tracking, 3D mapping & $30\sim150$ meters\\
		Long-range radar & Preceding vehicle detection & 150 meters \\
		\hline
	\end{tabular}
\end{table}

\subsection{Levels of perception}

\label{sec:lop}

In this section, we discuss how to categorize the sensing power of AVs with sensors listed in Table~\ref{tab:sensor} mounted. The Society of Automotive Engineers (SAE) has proposed a six-level classification criteria for autonomous vehicles \citep{avlevel}. L1 AVs can conduct adaptive cruise control, which is fulfilled by the long-range radar. From the perspective of traffic sensing, L1 AV can always detect the location and speed of its preceding vehicle in the same lane. From L2 to L5, AVs gradually take control from human drivers. To achieve that, AVs need to continuously observe the surrounding traffic conditions. From the perspective of traffic sensing, L2-L5 vehicles can detect or track the vehicles in their surrounding areas. Here we emphasize the difference between vehicle detection and vehicle tracking. Detection refers to the localization of a certain vehicle when it appears in the detection area of an AV, and tracking means that AV can keep track of a certain vehicle when it is within the detection area. To be precise, the task of detection does not require to ``memorize'' the detected vehicles in each time frame, while tracking requires the AV to keep track of the detected vehicles as long as they are within the detection range. Tracking is technically much more challenging than the detection. As of today, the detection technology is fairly mature, while the tracking technology is still not ready for real-world applications \citep{milan2016mot16}. The reason for the difference is that the detection/tracking is conducted frame by frame on AVs. If the AV processes 30 frame per-second, tracking requires to detect all the vehicles in each frame and match them correspondingly, while detection does not require to match the vehicles in different frames. The matching is challenging because vehicles often block each other, and this makes it difficult for machines to decide whether the detected vehicle is the same vehicle detected in previous frames. From the perspective of traffic sensing, detection only provides the locations of each vehicles but tracking can provide additional speed information. 

\begin{figure}[h]
	\centering
	\includegraphics[width=0.85\linewidth]{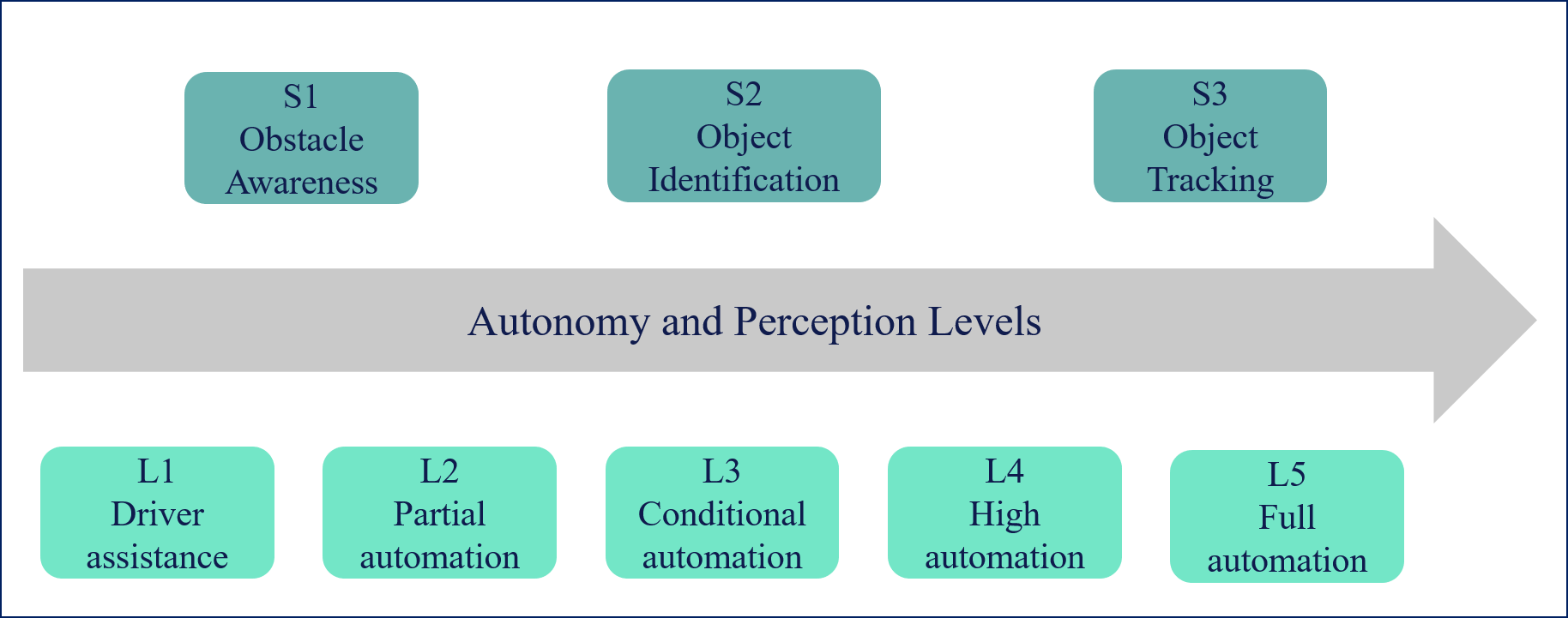}
	\caption{Overview of the perception levels.}
	\label{fig:percep}
\end{figure}

Analogous to the SAE's automation level definitions, we define three levels of sensing power for AVs, as presented in Figure~\ref{fig:percep}. 

The precise descriptions of the three perception levels are as follows.
\begin{itemize}
	\item $S_1$: The primary task for $S_1$ is to track the preceding vehicle, and this is originally used for the adaptive cruise control (ACC). However, the speed and location of the preceding vehicle are obtained for TSE.
	\item $S_2$: In addition to $S_1$, the primary task for $S_2$ is to detect and locate surrounding vehicles. Only vehicle counting is needed, and the speed information is not required in $S_2$.  
	\item $S_3$: In addition to $S_2$, the primary task for $S_3$ is to track every single vehicle in the detection area, hence the location and speed of each vehicle is monitored by AVs in $S_3$. 
\end{itemize}
Based on the definition of AV perception levels, $S_1$ requires a LRR dedicated for preceding vehicles, $S_2$ requires LiDAR/radar system, and $S_3$ require a comprehensive sensor fusion of camera, LiDAR, and radar. To be precise, section~\ref{sec:da} discusses how different sensors are combined to fulfill different levels of sensing power.

\subsection{Detection area of AVs}
\label{sec:da}
\begin{figure}[h]
	\centering
	\begin{subfigure}[b]{0.485\textwidth}
		\includegraphics[width=\textwidth]{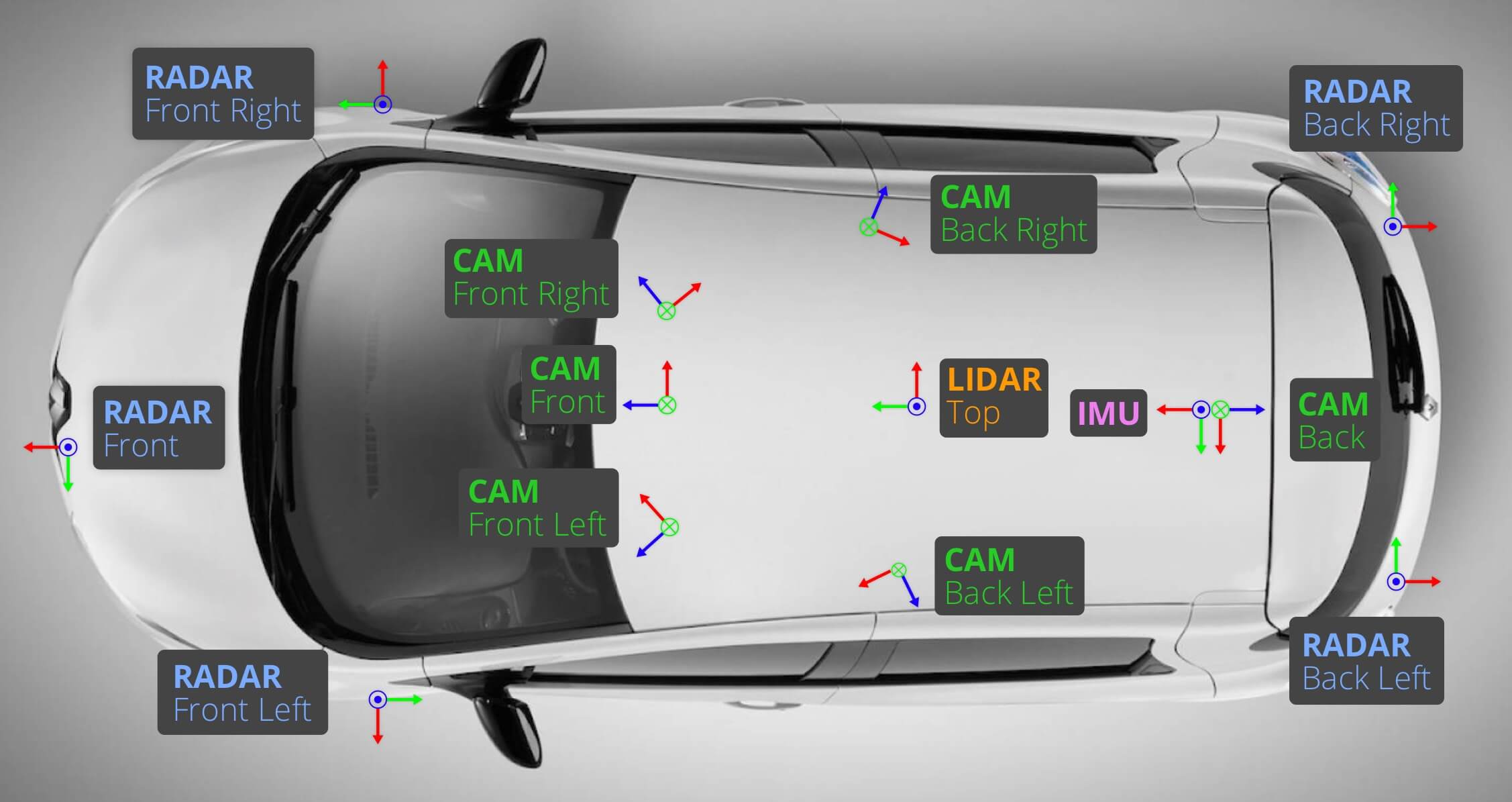}
		\caption{\footnotesize{nuScenes}}
		\label{fig:v1}
	\end{subfigure}
	\begin{subfigure}[b]{0.485\textwidth}
		\includegraphics[width=\textwidth]{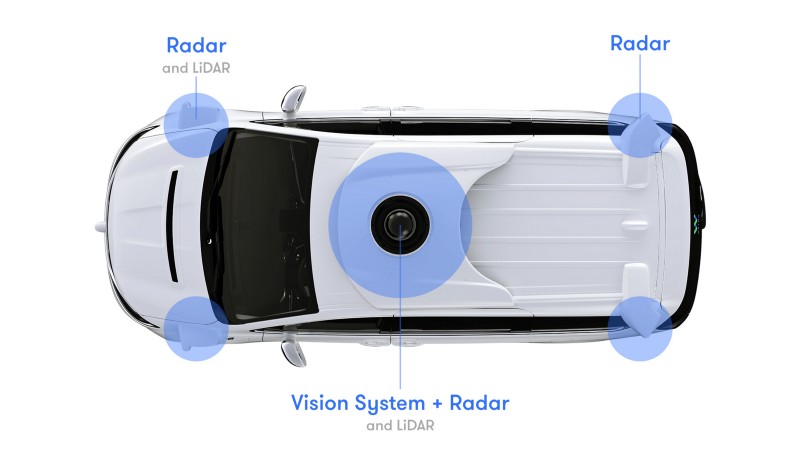}
		\caption{\footnotesize{Waymo}}
		\label{fig:v2}
	\end{subfigure}
	
	\caption{Two examples of sensor configurations, from \citet{nuscenes, waymo}.}
	\label{fig:sensor}
\end{figure}

We clearly define the surrounding area (or detection area) of AVs, which is used throughout the whole paper. The detection area of AVs depends on the sensor configurations. Figure~\ref{fig:sensor} presents two configurations of AV sensors. In the model of nuScenes, various sensors are mounted at different locations of an AV, while Waymo integrates most of the sensors on the top of the vehicle. Based on various sensor configurations on different AVs, the detection area of AVs can be different \citep{ihs}.

\begin{figure}[h]
	\centering
	\includegraphics[width=0.65\linewidth]{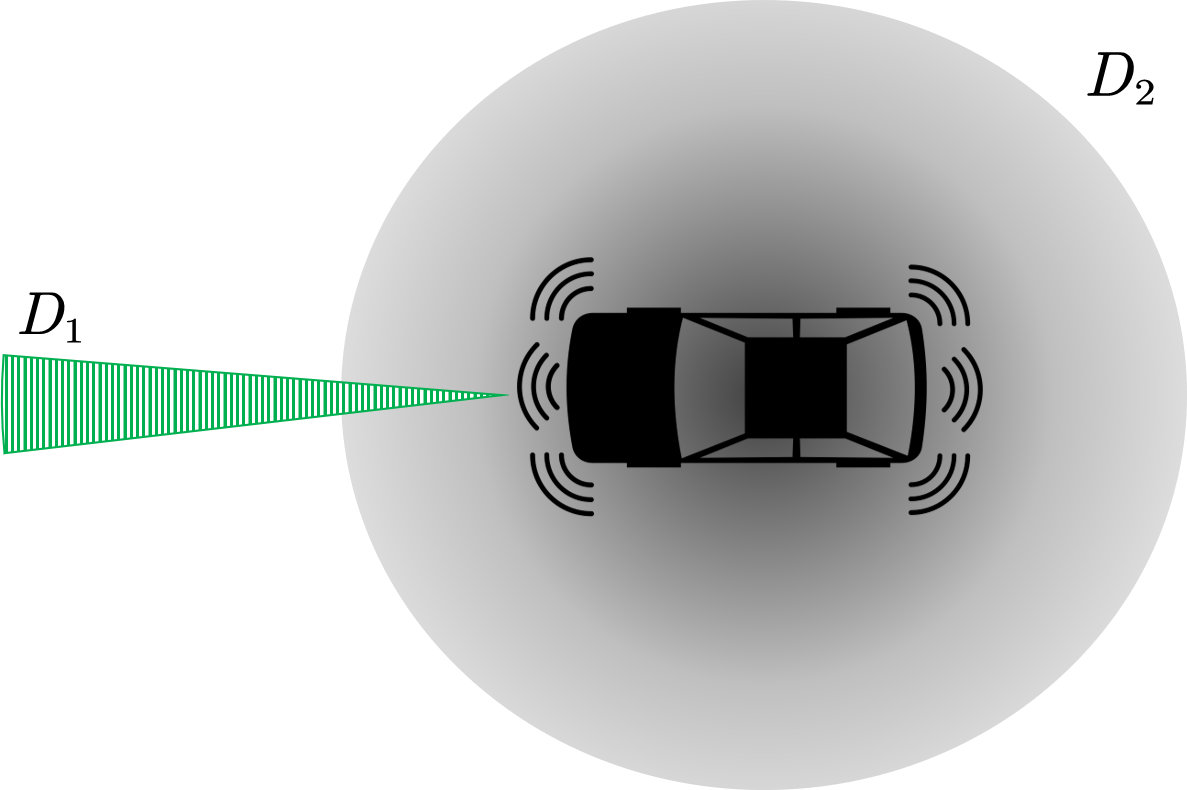}
	\caption{A simplified representation of AV detection area.}
	\label{fig:simp}
\end{figure}

In this paper, we adopt a simplified representation of AV detection area, as presented in Figure~\ref{fig:simp}.

The detection area in Figure~\ref{fig:simp} consists of two components: $D_1$ and $D_2$. $D_1$ is used for detecting the preceding vehicles, and it is fulfilled by the LRR; $D_2$ is for detecting the surrounding vehicles, which is supported by the combination of LiDAR and cameras. We assume only $D_1$ is active in $S_1$, while both $D_1$ and $D_2$ are active in $S_2$ and $S_3$, as presented in Table~\ref{tab:summary}.

\begin{table}[h]
	\caption{Summary of detection area and level of perceptions.}
	\centering
	\label{tab:summary}
	\begin{tabular}{|c|c|c|}
		\hline
		Sensing Power & Detection Area & Information Obtained  \\\hline\hline
		$S_1$ &  $D_1$& Speed/location of the preceding vehicle \\
		$S_2$ & $D_1$ and $D_2$ & Speed/location of the preceding vehicle, location of surround vehicles \\
		$S_3$ & $D_1$ and $D_2$ & Speed/location of the preceding vehicle and surrounding vehicles\\
		\hline
	\end{tabular}
\end{table}

\subsection{Data center}

\label{sec:datacenter}

\begin{figure}[h]
	\centering
	\includegraphics[width=0.75\linewidth]{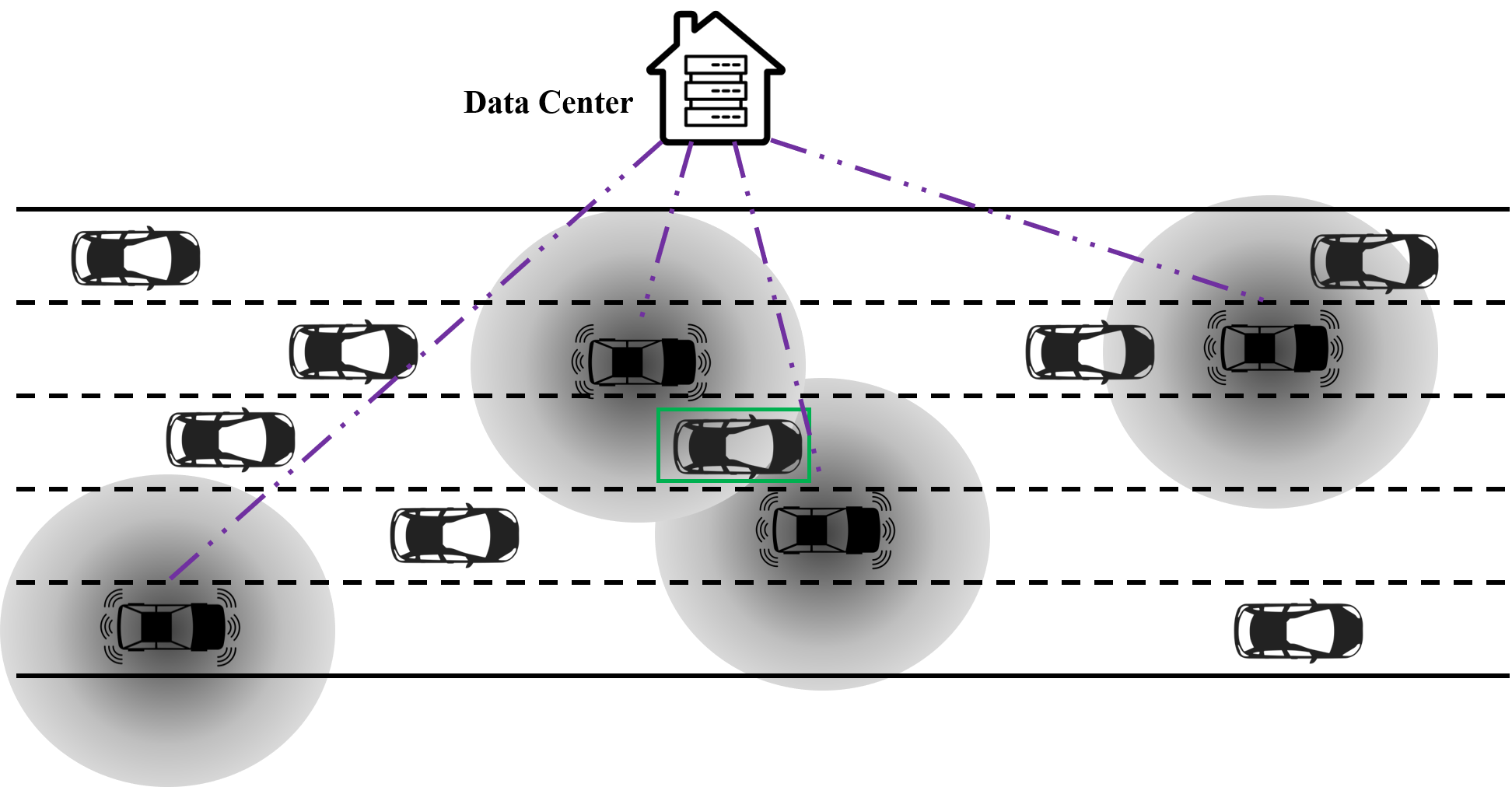}
	\caption{An illustration of the data center.}
	\label{fig:center}
\end{figure}

In this paper, we assume that there is a data center that receives all the information sent by AVs, as presented in  Figure~\ref{fig:center}. Due to the bandwidth and latency restrictions, AVs can not send all the raw data to the data center. Instead, the AV only sends the location and speed of the vehicles it has detected to the data center. The main task for the data center is to aggregate the information and remove the redundant information when the same vehicle is detected multiple times by different AVs. This task can be done by checking and matching the location of the detected vehicles. For example, the vehicle with green rectangle in Figure~\ref{fig:center} is detected by two AVs, hence two duplicate data points are sent to the data center and the data center is able to identify and clean these duplicate data points.  The localization accuracy is usually within the size of a standard vehicle, hence the accuracy for matching and cleaning is high \citep{wolcott2015fast}. In the numerical experiments, we will conduct sensitivity analysis to evaluate the impact of different matching accuracies.

\section{Formulation}
\label{sec:for}
In this section, we rigorously formulate the traffic state estimation (TSE) framework with AVs. We first present the notations, and then the traffic states variables are defined. A two-step estimation method is proposed: the first step directly translates the information observed by AVs and the second step employs data-driven methods to estimate the information that is not observed by AVs.

\subsection{Notations}
All the notations will be introduced in context, and Table~\ref{tab:notation} provides a summary of the commonly used notations for reference.
\begin{longtable}{p{3cm}p{10cm}}
	\caption{\footnotesize List of notations}
	\label{tab:notation}
	\endfirsthead
	\endhead	
	
	$l$ & Index of a certain lane \\
	$L$ & The set of all lane indices $l$\\
	$T$ & The set of all time points in the study period\\
	$T_h$ & The set of all time points in time interval $h$\\
	$x$ & A certain longitudinal location along the road\\
	$X_l$ & The set of all longitudinal locations on lane $l$\\
	$\star$ & Traffic state that is not directly observed by AVs\\
	$\mu(\cdot)$ & The Lebesgue measure for either one or two dimensional Euclidean space\\
	$|\cdot|$ & The counting measure for the countable sets\\
	\multicolumn{2}{c}{\textbf{Variables in a time-space region}}\\
	$h$ & Index of a certain time interval\\
	$H$ & The set of all indices $t$ in the study period\\	
	$s$ & Index of a certain longitudinal road segment\\
	$S$ & The set of all indices $s$\\
	$X_{ls}$ & The set of all longitudinal locations in road segment $s$ and lane $l$\\
	$A_l(h, s)$ & A cell in time-space region for time interval $h$ road segment $s$ and lane $l$\\
	$\bar{v}_l(h, s)$ & Average speed for time interval $h$ road segment $s$ and lane $l$ \\
	$\bar{q}_l(h, s)$ & Average traffic flow for  time interval $h$ road segment $s$ and lane $l$ \\
	$\bar{k}_l(h, s)$ & Average density for  time interval $h$ road segment $s$ and lane $l$ \\
	$a^i_l(h,s)$ & The headway area of vehicle $i$ in time-space region $A_l(h, s)$ \\
	
	\multicolumn{2}{c}{\textbf{Variables related to vehicles}}\\
	$i$ & Index of a certain vehicle\\
	$I$ & The set of all vehicle indices $i$\\
	$I_l(h,s)$ & The set of all vehicles indices in time interval $h$ road segment $s$ and lane $l$\\
	$v^i(t)$ & Instantaneous speed of vehicle $i$ at time $t$\\
	$h^i(t)$ & Instantaneous headway of vehicle $i$ at time $t$\\
	$x^i(t)$ & Instantaneous longitudinal location of vehicle $i$ at time $t$\\
	$l^i(t)$ & The lane in which vehicle $i$ is located at time $t$\\
	$\underline{t}^i$& The time point when the vehicle $i$ enters the road\\
	$\bar{t}^i$ & The time point when the vehicle $i$ exits the roads\\ 
	$d^i_l$ & The distance traveled by vehicle $i$ on lane $l$ \\
	$t^i_l$ & The time spent on lane $l$ by vehicle $i$\\
	\multicolumn{2}{c}{\textbf{Variables related to autonomous vehicles}}\\
	$j$ & Index of detection area\\
	$D_j$ & The detection area of an AV\\
	$I^A$ & The set of all autonomous vehicle indices\\
	$R^{i,D_j}_l(t,s)$ &  The detection area $D_j$ of all AVs in road segment $s$ on lane $l$\\
	$O_l^{D_j}$ & The set of time-space indices $(h,s)$ such that $X_{ls}$ is covered by the AV detection range in time interval $h$\\
	\multicolumn{2}{c}{\textbf{Variables related to the sensing framework}}\\
	$\tilde{k}_l(h,s)$ & The directly observed density for time interval $h$ road segment $s$ and lane $l$\\
	$\tilde{v}_l(h,s)$ & The directly observed speed for time interval $h$ road segment $s$ and lane $l$\\
	$\hat{k}_l(h,s)$ & The estimated density and speed for time interval $h$ road segment $s$ and lane $l$\\
	$\hat{v}_l(h,s)$ & The estimated speed for time interval $h$ road segment $s$ and lane $l$
\end{longtable}

\subsection{Modeling traffic states in time-space region}
\begin{figure}[h]
	\centering
	\includegraphics[width=0.75\linewidth]{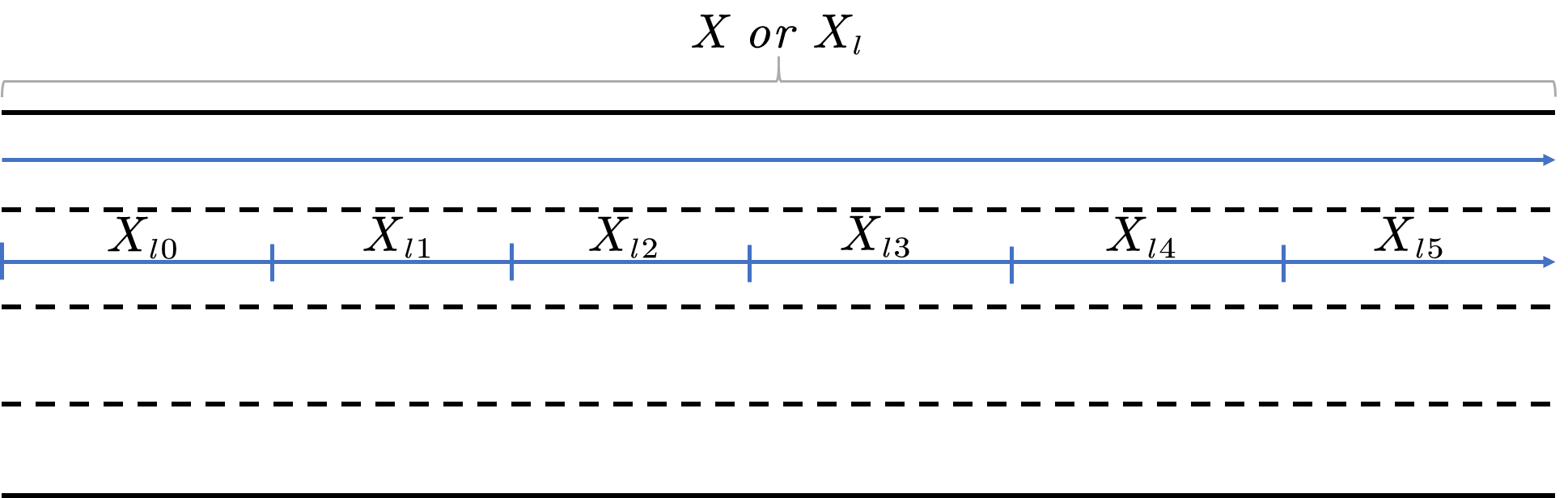}
	\caption{Overview of the highway discretization.}
	\label{fig:X}
\end{figure}

We consider a highway with $|L|$ lanes, where $L = \{0, 1, \cdots, |L|-1\}$. The operator $|\cdot|$ is the counting measure for countable sets. For each lane $l \in L$, we denote $X_{l}$ as the set of longitudinal locations on lane $l$, hence $\mu(X_{l})$ is the length of lane $l$. In this paper, we treat each lane as a one-dimensional line. Without loss of generality, we set the starting point of $X_{l}$ to be $0$, hence $X_l = [0, \mu(X_l)]$, where $\mu(X_l)$ is the length of lane $l$. Throughout the paper, we denote operator $\mu(\cdot)$ as the Lebesgue measure in either one or two dimensional Euclidean space, and it represents the length or area for one or two dimensional space. Note in this paper we assume the length of each lane is the same $\mu(X)= \mu(X_l), \forall l \in L$, while the proposed estimation method can be easily extended to accommodate different lane lengths.  We further discretize the road $X_l$ to $|S|$ equal road segments and each road segment is denoted by $X_{ls}$, where $s \in S$ is the index of the road segment and $S = \{0, 1, \cdots, |S|-1\}$. Hence we have $X_{ls} = [s \Delta_{X}, (s+1)\Delta_{X} ]$ and $\Delta_{X} = \frac{\mu(X)}{|S|}$. The above formulation is visualized in Figure~\ref{fig:X}.

We denote $i$ as the index of a certain vehicle and $I$ as the set of all vehicle indices. Suppose we know the location $\left( x^i(t), l^i(t) \right)$, speed $v^i(t)$, and space headway $h^i(t)$ of any vehicle $i$ at any time point $t \in T$, where $x^i(t)$ is the longitudinal location of vehicle $i$ at time $t$, $l^i(t)$ is the lane in which vehicle $i$ is located at time $t$, and $T$ is the set of all time points in the study period.
We assume each vehicle $i$ only enters the highway once. If a vehicle enters the highway multiple times, the vehicle at each entrance will be treated as a different vehicle. To obtain the traffic states, we construct the distance $d^i_l$ and time $t^i_l$ and headway area $\alpha^i_l$ from vehicle location $\left(x^i(t), l^i(t)\right)$, speed $v^i(t)$, and headway $h^i(t)$ for a certain vehicle $i$ based on \citet{edie1963discussion}. Suppose $\underline{t}^i$ denotes the time point when the vehicle enters the highway and $\bar{t}^i$ denotes the time point when the vehicle exits the highway, we denote the distance traveled and time spent on lane $l$ by vehicle $i$ as $d^i_l$ and $t^i_l$, respectively. Mathematically, $d^i_l$ and $t^i_l$ are presented in Equation~\ref{eq:dt}.
\begin{equation}
	\label{eq:dt}
	\begin{array}{llllll}
		d^i_l &=&  \mu \left( \left\{x^i(t) | \underline{t}^i \leq t \leq \bar{t}^i, l^i(t) = l  \right\}\right) \\ 
		t^i_l &=& \mu \left( \left\{t| \underline{t}^i \leq t \leq \bar{t}^i, l^i(t) = l  \right\}\right)
	\end{array}
\end{equation}

We use the headway area $\alpha^i_l$ to represent the headway between vehicle $i$ and its preceding vehicle on lane $l$ in the time-space region, and it is represented by Equation~\ref{eq:headway}. 
\begin{eqnarray}
	\label{eq:headway}
	\alpha^i_l  =\left\{ (t, x)| \underline{t}^i \leq t \leq \bar{t}^i, l^i(t) = l,  x^i(t) \leq x \leq x^i(t) + h^i(t)  \right\}
\end{eqnarray}


When we have the trajectories of all vehicles on the road, we can model the traffic states of each lane in a time-space region. Without loss of generality, we set the starting point of $t$ to be zero, hence we have $T = [0, \mu(T)]$, where $\mu(T)$ is the length of the study period. We discretize the study period $T$ to $|H|$ equal time intervals, where $H = \{0, 1, \cdots, |H|-1\}$. We denote $T_h$ as the set of time points for interval $h$, where $h \in H$.  Therefore, we have $T_h = [h \Delta_H, (h+1) \Delta_H]$, where $\Delta_H = \frac{\mu(T)}{|H|}$. In this paper, we use uniform discretization for $X_l$ and $T$ to simplify the formulation, while the proposed estimation methods work for arbitrary discretization scheme.

We use $A_{l}(h,s)$ to denote a certain cell in the time-space region for road segment $X_{ls}$ and time period $T_{h}$, as presented in Equation~\ref{eq:st}.

\begin{equation}
	\label{eq:st}
	\begin{array}{llllll}
		A_{l}(h,s) &=& T_h \otimes X_{ls} \\
		&=& \texttt{Polygon}\left( \left(h \Delta_H, s \Delta_{X}\right),
		\left( (h+1) \Delta_H, s \Delta_{X}\right), 
		\left( (h+1) \Delta_H, (s+1)\Delta_{X}\right),
		\left(h \Delta_H, (s+1)\Delta_{X}\right)  \right)\\
		&=& \left\{ (t, x)| h\Delta_H \leq t \leq (h+1) \Delta_H, s\Delta_{X} \leq x \leq (s+1) \Delta_{X}  \right\}\end{array} 
\end{equation}

We denote the headway area of vehicle $i$ in cell $A_{l}(h,s)$ by $a_l^i(h,s)$, as presented in Equation~\ref{eq:hv}.
\begin{eqnarray}
	\label{eq:hv}
	a_l^i(h,s) =  A_{l}(h,s) \cap \alpha^i_l
\end{eqnarray}

\begin{example}[Variable representation in time-space region]
	In this example, we illustrate the variables defined in the time-space region. We consider a one-lane road and the lane index is $l$. $X_l$ is segmented into $6$ road segments ($X_{l0}, \cdots, X_{l5}$), and $T$ is segmented into $10$ time intervals ($T_0, \cdots, T_9$), as presented in Figure~\ref{fig:ex}. The cell $A_l(0,4)$ is the intersection of $X_{l4}$ and $T_0$, $A_l(8,1)$ is the intersection of $X_{l1}$ and $T_8$.

	\begin{figure}[h]
		\centering
		\includegraphics[width=0.95\linewidth]{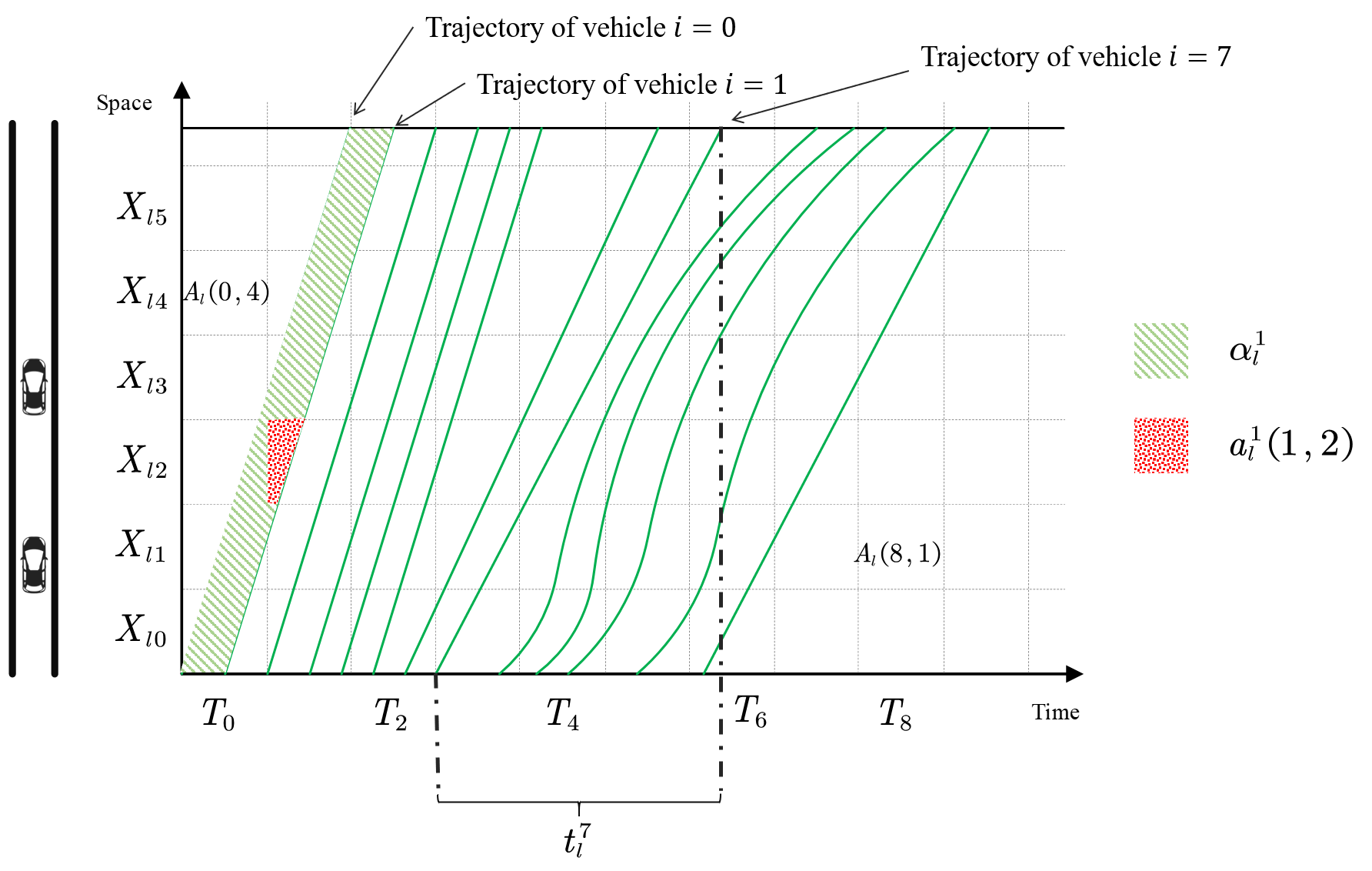}
		\caption{An example of variables in time-space region.}
		\label{fig:ex}
	\end{figure}

	Each green line in the time-space region represents the trajectory of a vehicle. In Figure~\ref{fig:ex}, we highlight the first ($i=0$), second ($i=1$) and the 8th ($i=7$) vehicle trajectory. The distance traveled by each vehicle is the same, hence $d_l^i = \mu(X_l) = \mu(X)$. We also highlight $t_l^7$ in Figure~\ref{fig:ex}, which represents the time spent by each vehicle $i=7$ on lane $l$.
	
	The headway area of vehicle $i=1$, denoted by $\alpha_l^1$, is represented by the green shaded area. The red shaded area, which represents $a_l^1(1,2)$, is the intersection of  $\alpha_l^1$ and $A_l(1,2)$ based on Equation~\ref{eq:hv}.
\end{example}

According to \citet{edie1963discussion, seo2015estimation}, we compute the traffic states variables, namely flow $\bar{q}_l(h,s)$, density $\bar{k}_l(h,s)$ and speed $\bar{v}_l(h,s)$, for each road segment $X_{ls}$ and time period $T_h$, as presented in Equation~\ref{eq:qvk}.

\begin{equation}
	\label{eq:qvk}
	\begin{array}{llllll}
		\bar{q}_l(h,s) &=& \frac{\sum_{i \in I} d^i_l }{ \sum_{i \in I} \mu(a_l^i(h,s))}\\
		\bar{k}_l(h,s) &=& \frac{\sum_{i \in I} t_l^i}{\sum_{i \in I} \mu(a_l^i(h,s))}\\
		\bar{v}_l(h,s) &=& \frac{\bar{q}_l(h,s)}{\bar{k}_l(h,s)} = \frac{\sum_{i \in I} d^i_l}{\sum_{i \in I} t_l^i}
	\end{array} 
\end{equation}

We treat the traffic states ({\em e.g.} flow $\bar{q}_l(h,s)$, density $\bar{k}_l(h,s)$ and speed $\bar{v}_l(h,s)$) estimated from full samples of vehicles $I$ as ground truth and unknown. 
In the following sections, we will develop a data-driven framework to estimate the traffic states from the partially observed traffic information obtained from autonomous vehicles under different levels of perception power.  

\subsection{Overview of the traffic sensing framework}

\begin{figure}[h]
	\centering
	\includegraphics[width=0.9\linewidth]{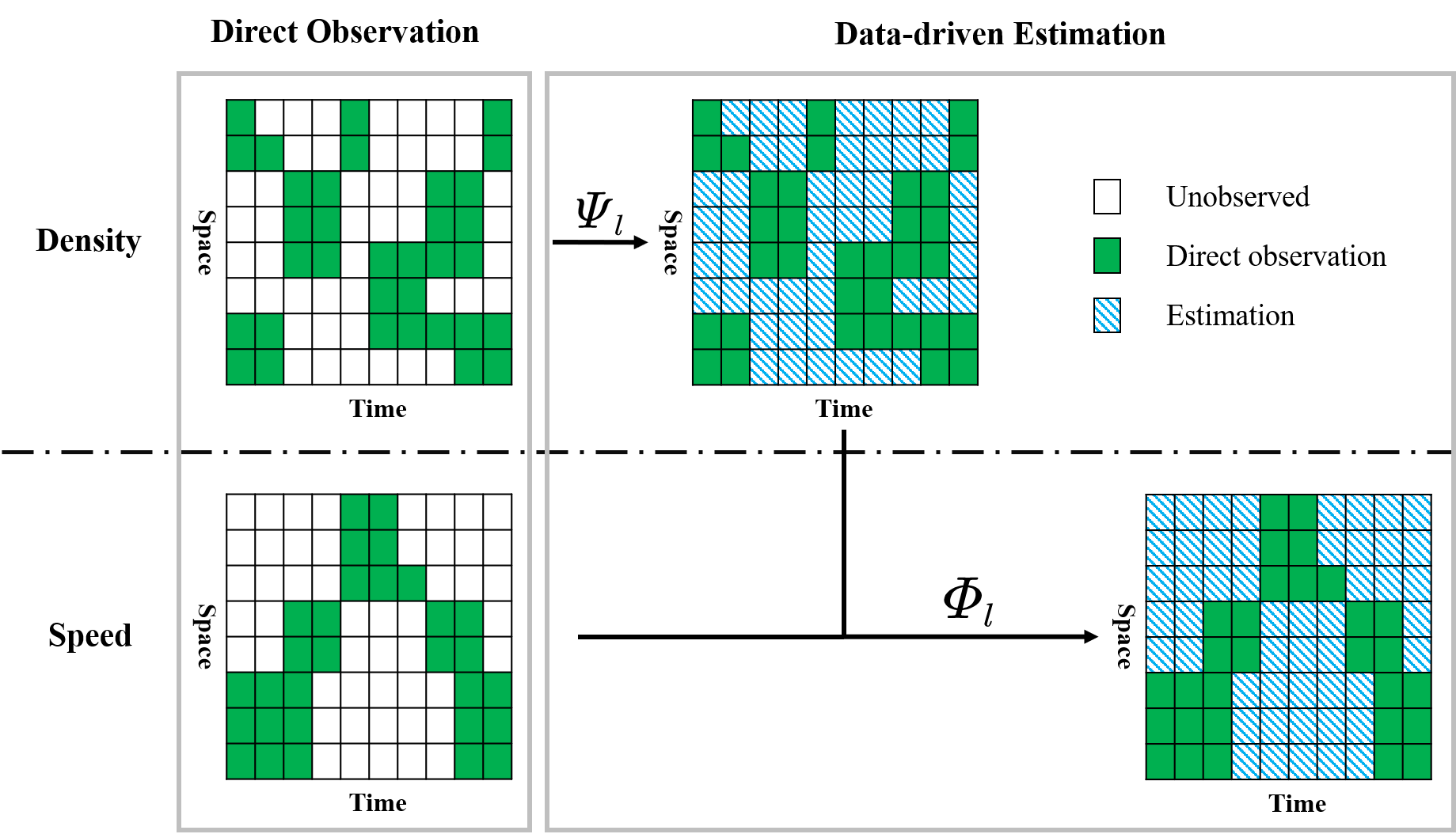}
	\caption{An overview of the traffic sensing framework.}
	\label{fig:frame}
\end{figure}

In this section, we present an overview of the traffic sensing framework. We assume a subset of vehicles are AVs, namely $I^A \subseteq I$, where $I^A$ denotes the index set of all AVs. The goal for the traffic sensing framework is to estimate the density and speed using the information observed by AVs.  Once the speed and density are estimated accurately, the traffic flow can be obtained by the conservation law \citep{bressan2015conservation}. The framework consists of two major parts: direct observation and data-driven estimation, as presented in Figure~\ref{fig:frame}. 

In the direct estimation step, density and speed are observed directly through AVs. Since AVs are moving observers \citep{wardrop1954method},  traffic states can only be observed partially for a certain set of time intervals and road segments ({\em i.e.} cells) in a time-space region. Section~\ref{sec:direct} will rigorously determine the set of cells that can be directly observed by AVs and compute the direct observations from information obtained by AVs. We will discuss the direct observation with different levels of sensing power. The second part aims at filling up the unobserved information with data-driven estimation methods. The functions $\Psi_l$ and $\Phi_l$ are used to estimate the unobserved density and speed on lane $l$, respectively. Details will be presented in section~\ref{sec:datadriven}.

\subsection{Direct observation}
\label{sec:direct}
In this section, we present to compute traffic states using the information that is directly observed by AVs under different levels of perception. Suppose the detection area of AV $i \in I^A$ at time $t$ is $R^i(t)$, and $R^i(t)$ consists of two parts, which are the detection area for preceding vehicles ($R^{i, D_1}$) and the detection area for surrounding vehicles ($R^{i, D_2}$), as discussed in section~\ref{sec:da}. We further denote the detection area $D_j$ of vehicle $i$ in road segment $s$ on lane $l$ by $R^{i, D_j}_l(t,s)$, as presented in Equation~\ref{eq:vrange}.

\begin{eqnarray}
	\label{eq:vrange}
	R^{i, D_j}_l(t,s) = R^{i, D_j}(t) \cap X_{ls}, \forall j\in \{1,2\}, i \in I^A
\end{eqnarray}

As discussed in section~\ref{sec:datacenter}, the detected information of all AVs will be aggregated by the data center. Hence the whole detection area by all AVs is denoted by $R_l^{D_j}(t,s)$, as presented in Equation~\ref{eq:allvrange}.

\begin{eqnarray}
	\label{eq:allvrange}
	R_l^{D_j}(t,s) = \cup_{i \in I^A} R^{i, D_j}_l(t,s), \forall j \in \{1,2\}
\end{eqnarray}

The next step is to discretize the detection area into the time-space region. We define $O_l^{D_j}$ as the set of time-space indices $(h,s)$ such that $X_{ls}$ is covered by the detection range in time interval $h$, as presented in Equation~\ref{eq:cov}.

\begin{eqnarray}
	\label{eq:cov}
	O_l^{D_j} = \{(h,s) | \exists t \in T_h,~\text{s.t.}~\mu(R_l^{D_j}(t,s)) \geq (1-\varepsilon) \mu(X_{ls}) \}, j = \{1, 2\}
\end{eqnarray}
where $\varepsilon$ is the tolerance and is set to $0.05$ in this paper.

Now we are ready to rigorously formulate the traffic states that can be directly observed by AVs under different levels of perception. As a notation convention, we use $\star$ to represent the information that cannot be directly observed by AVs, and $\tilde{k}_l, \tilde{v}_l$ denote the directly observed density and speed, respectively.

\subsubsection{$S_1$: tracking the preceding vehicle}

In the perception level $S_1$, an AV can only detect and track its preceding vehicle, and hence its detection area for density and speed is $O_l^{D_1}$. The observed density and speed can be represented in Equation~\ref{eq:os1}.

\begin{equation}
	\label{eq:os1}
	\begin{array}{lllllllll}
		\tilde{k}_l(h,s) &=& \begin{cases}
			\frac{\sum_{i \in I^A} t_l^i}{\sum_{i \in I^A} \mu(a_l^i(h,s))} & \text{if $(h,s) \in O_l^{D_1}$}\\
			\star & \text{else}
		\end{cases}
		\\\\
		\tilde{v}_l(h,s) &=& \begin{cases}
			\frac{\sum_{i \in I^A} d^i_l }{\sum_{i \in I^A} t_l^i} & \text{if $(h,s) \in O_l^{D_1}$}\\
			\star & \text{else}
		\end{cases}
	\end{array}
\end{equation}

Equation~\ref{eq:os1} is proven to be an accurate estimation of the traffic states \citep{seo2015estimation}. We note that some AVs also have a LRR mounted to track the following vehicle behind the AVs, and this situation can be accommodated by replacing the set $I^A$ with $I^A \cup \texttt{Following}(I^A)$ in Equation~\ref{eq:os1}, where $\texttt{Following}(I^A)$ represents all the vehicles that follow AVs in $I^A$.

When the AV market penetration rate is low, $O_l^{D_1}$ only covers a small fraction of all cells in the time-space region, especially for multi-lane highways. In contrast, $O_l^{D_2}$ covers more cells than $O_l^{D_2}$. Practically, it implies that the LiDAR and cameras are the major sensors for traffic sensing.

\subsubsection{$S_2$: locating surrounding vehicles}
In the perception level $S_2$, both $D_1$ and $D_2$ are enabled by the LRR, LiDAR and cameras, while $D_2$ can only detect the location of surrounding vehicles. Hence the density can be observed in both $D_1$ and $D_2$, and the speed is only observed in $D_1$. The estimation method for $D_1$ cannot be used for $D_2$ since the preceding vehicles of the detected vehicle might not be detected, hence $\alpha_l^i(h,s)$ cannot be estimated accurately. Instead, $D_2$ provides a snapshot of the traffic density at a certain time point, and we can compute the density of time interval $h$ by taking the average of all snapshots, as presented in Equation~\ref{eq:os2}.

\begin{equation}
	\label{eq:os2}
	\begin{array}{lllllllllll}
		\tilde{k}_l(h,s) &=& 
		\begin{cases}
			\frac{\sum_{i \in A} t_l^i}{\sum_{i \in A} \mu(a_l^i(h,s))} & \text{if $(h,s) \in O_l^{D_1}$}\\
			\frac{1}{\mu(T_h^o(l,s))}\int_{t \in T_h^o(l,s)}    \frac{|I_{l}^o(t,s)|}{\mu(X_{ls})} dt & \text{else if $(h,s) \in O_l^{D_2}$} \\
			\star & \text{else}\\
		\end{cases}\\\\
		\tilde{v}_l(h,s) &=& 
		\begin{cases}
			\frac{\sum_{i \in A} d^i_l }{\sum_{i \in A} t_l^i} & \text{if $(h,s) \in O_l^{D_1}$}\\
			\star & \text{else}
		\end{cases}
	\end{array}
\end{equation}
where $T_{h}^o(l,s) = \{t \in T_h| R_l^{D_2}(t,s) \geq (1-\varepsilon)|X_{ls}| \}$ represents the set of time indices when $X_{ls}$ is covered by the $D_2$ in $T_h$, and $I_l^o(t,s) = \{i \in I|  x^i(t) \in R_l^{D_2}(t,s) \}$ represents the set of vehicles detected by $D_2$ in $X_{ls}$ at time $t$.

\subsubsection{$S_3$: tracking surrounding vehicles}
In the perception level $S_3$, both localization and tracking are enabled by the LRR, LiDAR and cameras. In addition to the information obtained by $S_2$, speed information of surrounding vehicles in $D_2$ are also available. Similar to the density estimation, we first computed the instantaneous speed of a cell at a certain time point by taking the harmonic mean of all detected vehicles, and then the average speed of a cell is computed by taking the average of all time points, as presented by Equation~\ref{eq:os3}.

\begin{equation}
	\label{eq:os3}
	\begin{array}{lllllllllllll}
		\tilde{k}_l(h,s) &=& 
		\begin{cases}
			\frac{\sum_{i \in A} t_l^i}{\sum_{i \in A} \mu(a_l^i(h,s))} & \text{if $(h,s) \in O_l^{D_1}$}\\
			\frac{1}{\mu(T_h^o(l,s))}\int_{t \in T_h^o(l,s)} \frac{ |I_{l}(t,s)|}{\mu(X_{ls}|} dt & \text{else if $(h,s) \in O_l^{D_2}$} \\
			\star & \text{else}\\
		\end{cases}\\\\
		\tilde{v}_l(h,s) &=& \begin{cases}
			\frac{\sum_{i \in A} d^i_l }{\sum_{i \in A} t_l^i} & \text{if $(h,s) \in O_l^{D_1}$}\\
			\frac{1}{\mu(T_h^o(l,s))}\int_{t \in T_h^o(l,s)} \texttt{hmean} \left( \left\{ v^i(t) | x^i(t) \in X_{ls}, i \in  I \right\}   \right) dt & \text{else if $(h,s)\in O_l^{D_2}$}\\
			\star & \text{else}
		\end{cases}
	\end{array}
\end{equation}
where $\texttt{hmean}(\cdot)$ represents the harmonic mean. Though $S_3$ provides the most speed information, the directly observed density is the same for $S_2$ and $S_3$. Overall, the sensing power of AV increases as more cells are directly observed from $S_1$ to $S_3$. In the following section, we will present to fill the $\star$ using data-driven methods.

\subsection{Data-driven estimation method}
\label{sec:datadriven}

In this section, we propose a data-driven framework to estimate the unobserved density and speed in $\tilde{k}_l(h,s), \tilde{v}_l(h,s)$. To differentiate the density (speed) before and after the estimation, we use $\hat{k}_l(h,s)$ and $\hat{v}_l(h,s)$ to represent the estimated density and speed for time interval $h$ road segment $s$ and lane $l$, while $\tilde{k}_l(h,s), \tilde{v}_l(h,s)$ denote the density and speed before the estimation ({\em i.e.} after the direct observation). The method consists of two steps: 1) estimate the unobserved density $\hat{k}_l(h,s)$ given the observed density $\tilde{k}_l(h,s)$; 2) estimate the unobserved speed $\hat{v}_l(h,s)$ given that the density $\hat{k}_l(h,s)$ is fully known from estimation and speed $\tilde{v}_l(h,s)$ is partially known from direct observation.

We present the generalized form for estimating the unobserved density and speed in Equation~\ref{eq:ek} and Equation~\ref{eq:ev}, respectively.

\begin{eqnarray}
	\label{eq:ek}
	\hat{k}_l(h,s) &=& \Psi_l\left(h, s, \{\tilde{k}_{l'}\}_{l'}\right)\\
	\label{eq:ev}
	\hat{v}_l(h,s) &=& \Phi_l\left(h, s, \{\tilde{v}_{l'}\}_{l'},  \{\hat{k}_{l'}\}_{l'} \right)
\end{eqnarray}
where $\Psi_l$ is a generalized function that takes the observed density $\{\tilde{k}_l\}_l$ and time/space index $h,s$ as input and outputs the estimated density.  $\Phi_l$ is also a generalized function to estimate speed, while its inputs include the observed speed  $\{\tilde{v}_l\}_l$, the estimated density $\{\hat{k}_l\}_l$, and the time/space index $h,s$.
In this paper, we propose matrix completion-based methods for $\Psi_l$, and both matrix completion-based  and regression-based methods for $\Phi_l$. Details are presented in the following subsections.

\subsubsection{Matrix completion-based methods}

The matrix completion-based model can be used to estimate either density or speed. We first assume that densities (speeds) in certain cells are directly observed by the AVs, as presented in Equation~\ref{eq:vd}.
\begin{equation}
	\label{eq:vd}
	\begin{array}{llllllllll}
		\hat{k}_l(h,s) &=& \tilde{k}_l(h,s), \forall (h,s) \in O_l^{k, S_p}, p \in \{1,2,3\}\\
		\hat{v}_l(h,s)  &=& \tilde{v}_l(h,s), \forall (h,s) \in O_l^{v, S_p}, p \in \{1,2,3\}
	\end{array}
\end{equation}
where we denote $O_l^{k, S_p}$ and $O_l^{v, S_p}$ as the detection area for density and speed of lane $l$ in time-space region with sensing power $p \in \{ 1,2,3\}$. Precisely, $O_l^{k, S_1} = O_l^{D_1}, O_l^{v, S_1} = O_l^{D_1}, O_l^{k, S_2} = O_l^{D_1} \cup O_l^{D_2}, O_l^{v, S_2} = O_l^{D_1}$ and $O_l^{k, S_3} = O_l^{D_1} \cup O_l^{D_2}, O_l^{v, S_3} = O_l^{D_1} \cup O_l^{D_2}$.

For each lane $l$, the estimated density $\hat{k}_l$ (or speed $\hat{v}_l$) forms a matrix in the time-space region, and each row represents a certain road segment $s$ and each column represents a certain time interval $h$. Some entries ($(h,s) \notin O_l^{k, S_p}$ or $(h,s) \notin O_l^{v, S_p}$) in the density matrix (or speed matrix) are missing. To fill the missing entries, many standard matrix completion methods can be used. For example, the naive imputation (imputing with the average values across all time intervals or across all cells), k-nearest neighbor (k-NN) imputation \citep{troyanskaya2001missing}, and
the singular-value decomposition (SVD)-based \textsc{SoftImpute} algorithm \citep{hastie2015matrix}.

\subsubsection{Regression-based methods}

The speed data can also be estimated by a regression-based model given the density $\hat{k}_l(h,s)$ is fully estimated. We train a regression model $f_l$ to estimate the speed from densities for lane $l$, as presented in Equation~\ref{eq:rg}.

\begin{eqnarray}
	\label{eq:rg}
	\hat{v}_l(h,s) = f_l\left (\right\{\hat{k}_{l'}(h',s') :  l - \delta_l \leq l' \leq l + \delta_l, h-\delta_h \leq h' \leq h,s - \delta_s \leq  s' \leq  s + \delta_s, l' \in L, s \in S, h \in H  \left\} \right)
\end{eqnarray}
where $\delta_l, \delta_h, \delta_s$ represent the number of nearby lanes, time intervals and road segments considered in the regression model. The intuition behind the regression model is that the speed of a cell can be inferred by the densities of its neighboring cells. A specific example of $f_l$ is the fundamental digram \citep{newell1993simplified}, which is formulated by $\hat{v}_l(h,s) = f_l\left(\hat{k}_l(h,s)\right)$ by setting  $\delta_l = \delta_h = \delta_s = 0$. 

\begin{figure}[h]
	\centering
	\includegraphics[width=0.9\linewidth]{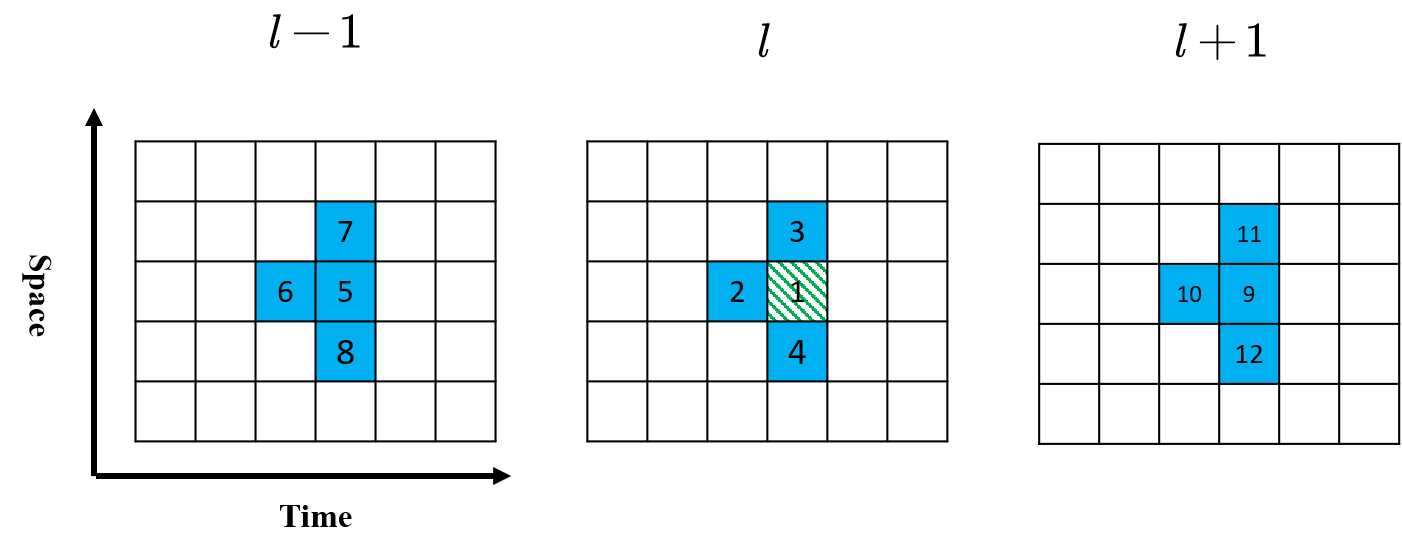}
	\caption{Cells used for speed estimation.}
	\label{fig:regression}
\end{figure}

In this paper, we adopt a simplified function $f_l(\cdot)$ presented in Figure~\ref{fig:regression}. Suppose we want to estimate  the speed for cell 1, there are 12 neighboring cells (including cell 1) considered as inputs. The regression methods adopted in this paper are Lasso \citep{tibshirani1996regression} and random forests \citep{breiman2001random}. 

\section{Solution Algorithms}
\label{sec:sol}
In this section, we discuss some practical issues regarding the traffic sensing framework proposed in section~\ref{sec:for}.

\subsection{Computation of $a_l^{i}(h,s)$}
To obtain the ground truth (Equation~\ref{eq:qvk}) and the observed density (Equation~\ref{eq:os1}), $a_l^i(h,s)$, which denotes the headway area of vehicle $i$ in cell $A_{l}(h,s)$ (Equation~\ref{eq:hv}), needs to be computed in the time-space region. $a_l^i(h,s)$ is computed by intersecting $A_{l}(h,s)$ and $\alpha^i_l$, and $A_{l}(h,s)$ can be represented by a rectangle in the time-space region. The headway area for vehicle $i$ $\alpha^i_l$ is usually banded \citep{seo2015estimation}, which can be approximated by a polygon. Therefore,  $a_l^i(h,s)$ can also be represented by a polygon, and the interaction of a rectangle (which is also a special polygon) and a polygon can be conducted efficiently \citep{strobl2017dimensionally}.

\subsection{Sampling rate}
As discussed in section~\ref{sec:datacenter}, the AVs send messages to data center periodically. Let the sampling rate denote the message sending frequency, and we assume that all AVs have the same sampling rate. When the sampling rate is high, the data center can obtain the density and speed information in high temporal resolution, hence the traffic sensing can be accurate. On the other hand, the sampling rate is limited by the bandwidth and latency of the message transmission network. In the numerical experiments, sensitivity analysis will be conducted to study the impact of sampling rate.

\subsection{Cross-validation}

In the data-driven method presented in section~\ref{sec:datadriven}, the cross-validation is conducted for model selection in both matrix completion-based and regression-based methods. 

In the matrix completion-based model, we use cross-validation to select the maximal rank in the \textsc{SoftImpute} and the number of nearest neighbors in the k-NN imputation \citep{kanagal2010rank}. To perform the cross-validation for the matrix completion,  we randomly hide $10\%$ of the matrix entries and run the imputation methods on the rest of entries. Then we measure the imputation accuracy by comparing the imputed values and the actual values on the $10\%$ hidden entries .

In the regression-based model, 5-fold cross-validation is performed to select the optimal parameter settings for different regression methods, such as the weight of regularization term in Lasso, number of base estimators in random forests.

\section{Numerical Experiments}
\label{sec:exp}
In this section, we conduct the numerical experiments with NGSIM data to examine the proposed TSE framework. All the experiments below are conducted on a desktop with Intel Core i7-6700K CPU @ 4.00GHz $\times$ 8, 2133 MHz 2 $\times$ 16GB RAM, 500GB SSD, and the programming language is Python 3.6.8.

\subsection{Data and experiment setups}
\begin{figure}[h]
	\centering
	\includegraphics[width=0.85\linewidth]{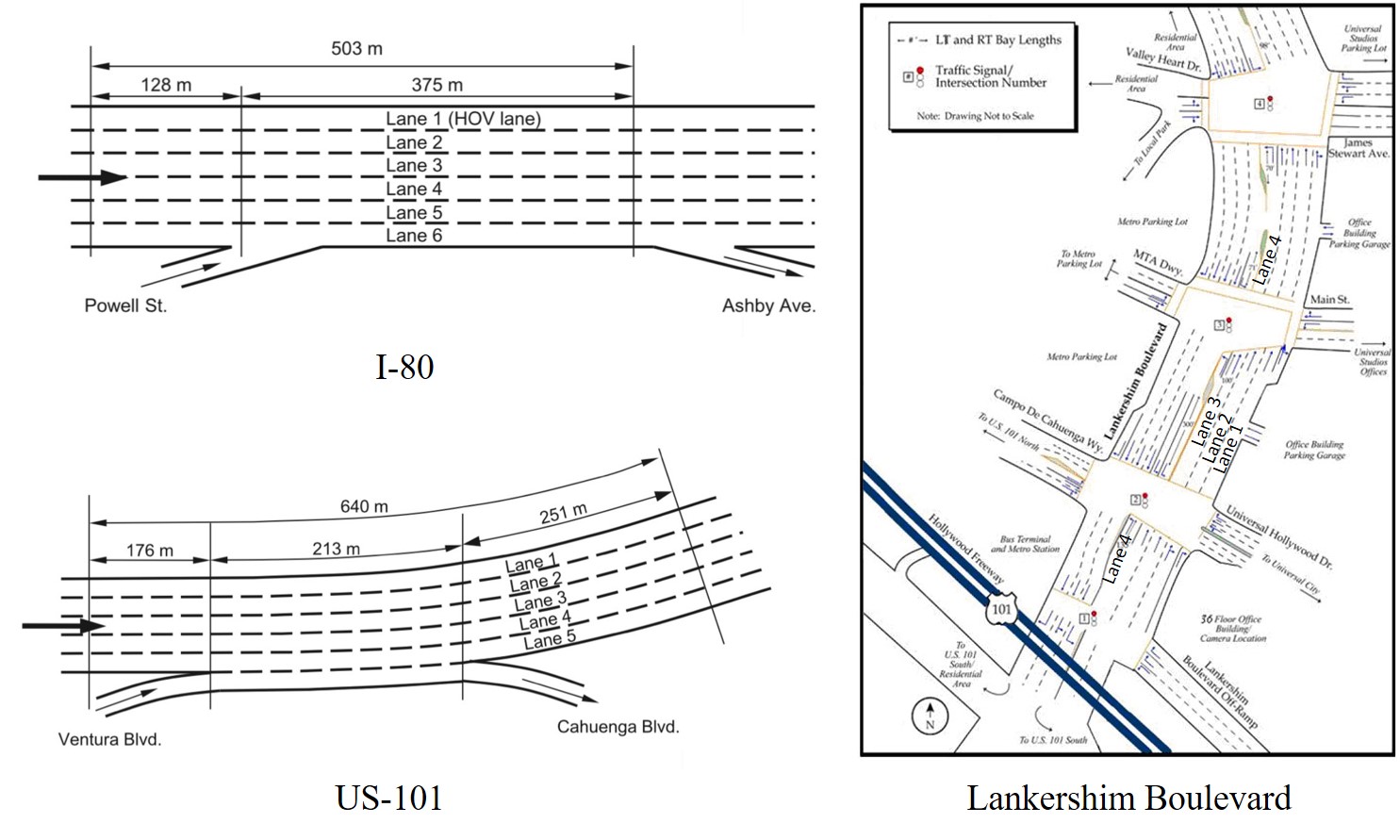}
	\caption{Overview of three networks \footnotesize{(adapted from NGSIM website \citep{alexiadis2004next} and \citet{he2017research})}.}
	\label{fig:roads}
\end{figure}

We use the Next Generation Simulation (NGSIM) data to validate the proposed framework. NGSIM data contains high-resolution vehicle trajectory data on different roads \citep{alexiadis2004next}. Our experiments are conducted on I-80, US-101 and Lankershim Boulevard, and the overviews of the three roads are presented in Figure~\ref{fig:roads}. NGSIM data is collected using digital video camera, and its temporal resolution is 100ms. Details of the three roads can be found in \citet{alexiadis2004next, he2017research}.

We assume that a random set of vehicles are AVs and the AVs can perceive the surrounding traffic conditions. Given the limited information collected by AVs, we estimate the traffic states using the proposed framework. We further compare the estimation results with the ground truth computed from the full vehicle trajectory data. The Normalized Root Mean Squared Error (NRMSE), Symmetric Mean Absolute Percentage Error (SMAPE1, SMAPE2) will be used to examine the estimation accuracy, as presented by Equation~\ref{eq:measure}. SMAPE2 is considered as a robust version of SMAPE1 \citep{li2014multimodel}.

\begin{equation}
	\setstretch{2.25}
	\begin{array}{lllllllll}
		\text{NRMSE}(z, \hat{z}) = \sqrt{\frac{\sum_{\mu \in \Mu} (z_\mu - \hat{z}_\mu)^2 }{\sum_{\mu \in \Mu} z_\mu^2}}\\
		\text{SMAPE1} (z, \hat{z}) =  \frac{1}{|\Mu|}\sum_{\mu \in \Mu} \frac{|z_{\mu} - \hat{z}_\mu |}{z_{\mu} + \hat{z}_\mu}\\
		\text{SMAPE2}(z, \hat{z}) = \frac{ \sum_{\mu \in \Mu} |z_{\mu} - \hat{z}_\mu |}{ \sum_{\mu \in \Mu} (z_{\mu} + \hat{z}_\mu)}
	\end{array}
	\label{eq:measure}
\end{equation}
where $z$ is the true vector, $\hat{z}$ is the estimated vector, $\mu$ is the index of the vector, and $M$ is the set of indices in vector $z$ and $\hat{z}$. When comparing two matrices, we flatten the matrices to vector and then conduct the comparison.

Here we describe all the factors that affect the estimation results. The market penetration rate denotes the proportions of AVs in the fleet. In the experiments, we assume the AVs are uniformly distributed in the fleet. The detection area $D_1$ is a ray fulfilled by LRR and $D_2$ is a circle fulfilled by LiDAR. We assume LiDAR has a detection range and it might also oversee a vehicle with a certain probability (referred to as missing rate). The AVs can be at one level of perception, as discussed in section~\ref{sec:sensing}. The sampling rate of data center can be different. In addition, different data-driven estimation methods are used to estimate the density and speed, as presented in section~\ref{sec:datadriven}. We define LR1 and LR2 as Lasso regression, and RF1 and RF2 as random forests regression. The number \textsc{1} means only cells $1$ to $4$ are used as inputs, while the number \textsc{2} means all $12$ cells in Figure~\ref{fig:regression} are used as inputs. SI denotes the \textsc{SoftImpute}, KNN denotes the k-nearest neighbor imputation, and NI denotes the naive imputation by simply replacing missing entries with the mean of each column.

{\bf Baseline setting}: The market penetration rate of AVs is $5\%$. The detection range of LRR is $150$ meters, and the detection range of LiDAR is $50$ meters with $5\%$ missing rate. The level of perception is $S_3$, and the speed is detected without any noise. The sampling rate of data center is 1 Hz. SI is used to estimate density and LR2 is used to estimate speed. We set $|H| = 90$ and $|S| = 60$.

\subsection{Basic results} We first run the proposed estimation method with the baseline setting. The estimation method takes around 7 minutes to estimate all three roads, and the most time consuming part is the information aggregation in the data center (discussed in section~\ref{sec:datacenter}) and the computation of Equation~\ref{eq:os3}. The estimation accuracy is computed by averaging the NRMSE, SMAPE1 and SPAME2 through all lanes, and the results are presented in Figure~\ref{tab:basic}. 

\begin{table}[h]
	\centering
	\begin{tabular}{lllllll}
		\toprule
		\multirow{2}{*}{Measures} & \multicolumn{3}{c}{Density}  & \multicolumn{3}{c}{Speed}\\
		\cmidrule(r){2-7}
		~    & NRMSE & SMAPE1 &  SMAPE2 & NRMSE & SMAPE1 &  SMAPE2 \\
		\midrule
		I-80 & 18.61 & 7.65 & 6.87 & 9.40 & 3.17 & 2.73\\
		US-101 & 18.28 & 7.76 & 6.89 & 7.49 & 2.88& 2.40\\
		Lankershim & 50.94 & 22.73 & 19.71 & 24.08 & 10.01& 8.00\\
		\bottomrule
	\end{tabular}
	\vspace{0.5em}
	\caption{Estimation accuracy with basic setting.}
	\label{tab:basic}
\end{table}

In general, the estimation method yields accurate estimation on highways (I-80 and US-101), while it underperforms on the complex arterial road (Lankershim Boulevard). Estimation accuracy of speed is always higher than that of density, which is because the density estimation requires every vehicle being sensed while speed estimation only needs a small fraction of vehicles being sensed \citep{long2002probe}.

{\bf Estimation accuracy on different lanes.} We then examine the performance of the proposed method on each lane separately, and the estimation accuracy of each lane is summarized in Table~\ref{tab:basiclane}. 

\begin{table}[h]
	\centering
	\begin{tabular}{lllllllllll}
		\toprule
		\multirow{2}{*}{Item} &\multirow{2}{*}{Lanes}  & \multicolumn{6}{c}{I-80}&  \\
		\cmidrule(r){3-8}
		~&~    & 1 & 2 &  3 & 4 & 5 &  6  \\
		\midrule
		\multirow{3}{*}{Density} & NRMSE &32.08 & 15.24 & 18.08 & 15.46 & 15.51 & 15.30\\
		&SMAPE1 & 13.43 & 6.21 & 7.17 & 6.02 & 6.45& 6.61\\
		&SMAPE2  & 12.40 & 5.57 & 6.39 & 5.35 & 5.71& 5.75\\
		\midrule
		\multirow{3}{*}{Speed} & NRMSE & 6.82 & 9.68 & 9.17 & 10.99 & 9.99 & 9.73\\
		&SMAPE1 & 1.93 & 3.71 & 3.17 & 3.77 & 3.26& 3.18\\
		&SMAPE2 & 1.94 & 3.01 & 2.73 & 3.17 & 2.83& 2.73\\
		\midrule
		\multirow{2}{*}{Item} & \multirow{2}{*}{Lanes}  & \multicolumn{5}{c}{US-101}& \multicolumn{4}{c}{Lankershim Blvd} \\
		\cmidrule(r){3-11}
		&~    & 1 & 2 &  3 & 4 & 5 &  1 & 2 & 3 & 4  \\
		\midrule
		\multirow{3}{*}{Density} & NRMSE & 17.90 & 17.95 & 18.47 & 18.09 & 18.96 & 45.49& 41.30&43.75&73.19\\
		&SMAPE1 & 7.56 & 7.62 & 7.60 & 7.75 & 8.27& 20.92 &21.00 &21.37 &27.63\\
		&SMAPE2 & 6.75 & 6.79 & 6.83 & 6.79& 7.28& 17.53 & 16.59&16.95 &27.73\\
		\midrule
		\multirow{3}{*}{Speed} & NRMSE &8.76 & 7.85 & 6.86 & 7.18 & 6.76 & 23.94 & 21.33& 25.26&25.78\\
		&SMAPE1 & 3.62 & 3.05 & 2.65 & 2.56 & 2.50& 10.22 & 8.77& 10.97& 10.08\\
		&SMAPE2 & 2.87 & 2.48 & 2.22 & 2.19 & 2.21& 7.71 & 6.77&8.49 &9.02 \\
		\bottomrule
	\end{tabular}
	\vspace{0.5em}
	\caption{Estimation accuracy on each lane with basic setting.}
	\label{tab:basiclane}
\end{table}

On can read from Table~\ref{tab:basiclane} that the proposed method performs similarly on most lanes. One interesting observation is that the proposed method performs well on Lane 6 on I-80, Lane 5 on US-101 and Lane 1 on Lankershim Blvd, and those lanes are merged with ramps. This implies that the proposed method has potentials to work well on merging intersections.

The proposed method performs differently on lanes that are near the edge of roads. For example, the proposed method yields the worst density estimation and the best speed estimation on lane 1 of I-80, which is an HOV lane. The vehicle headway is relatively large on the HOV lane, hence estimating density is more challenging given limited detection range of LiDAR. In contrast, speed on HOV lane is relatively stable, making the speed estimation easy. In addition, the estimation accuracy of the Lane 4 of Lankershim Blvd is low, as a result of the physical discontinuity of the lane.

\begin{figure}[h]
	\centering
	\includegraphics[width=0.95\linewidth]{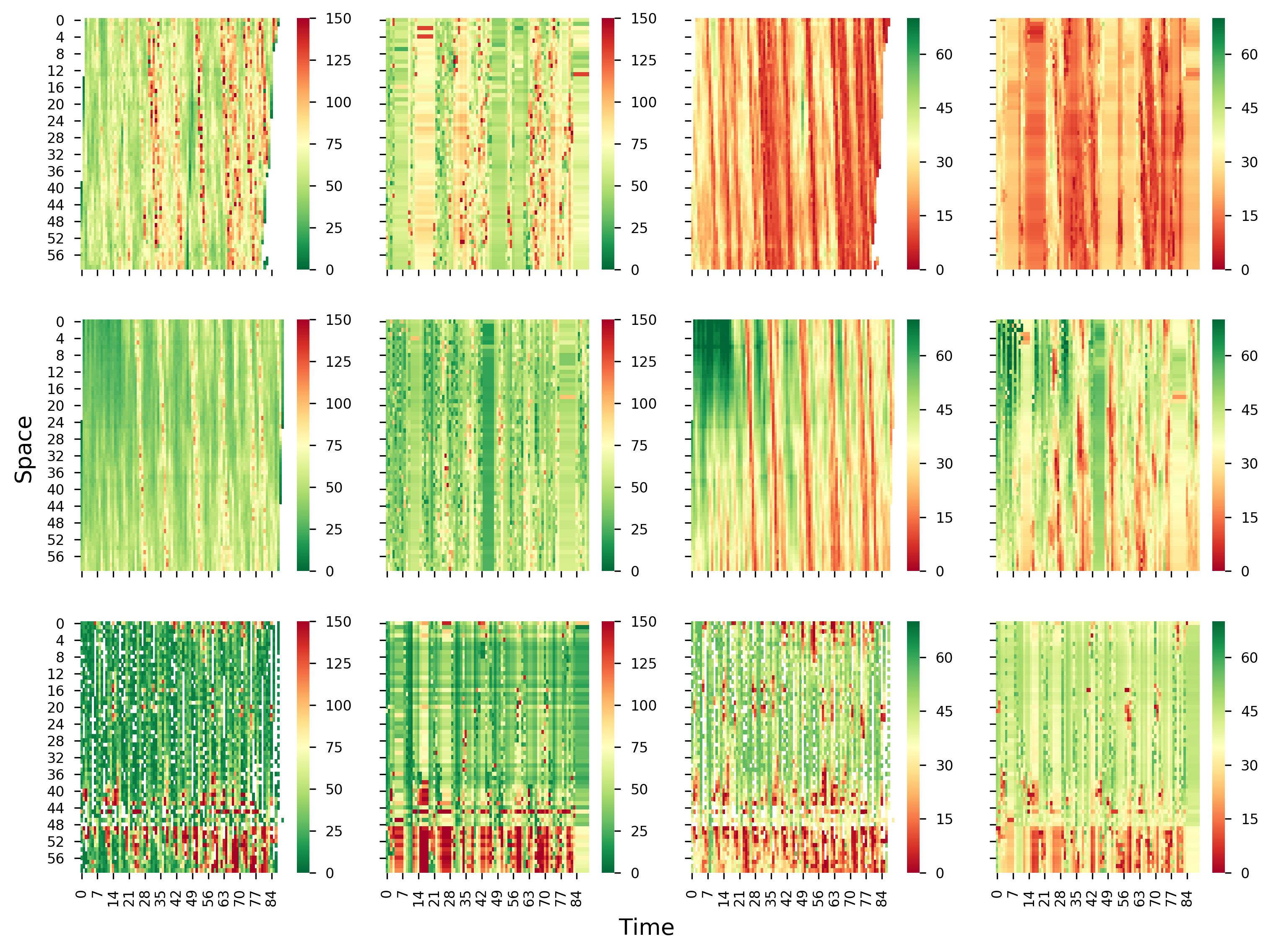}
	\caption{True and estimated density and speed for lane 2 \footnotesize{(first row: I-80, second row: US-101. third row: Lankershim Blvd; first column: ground truth density, second column: estimated density, third column: ground truth speed, fourth column: estimated speed; density unit: veh/km, speed unit: km/hour)}.}
	\label{fig:basic2}
\end{figure}

\begin{figure}[h]
	\centering
	\includegraphics[width=0.95\linewidth]{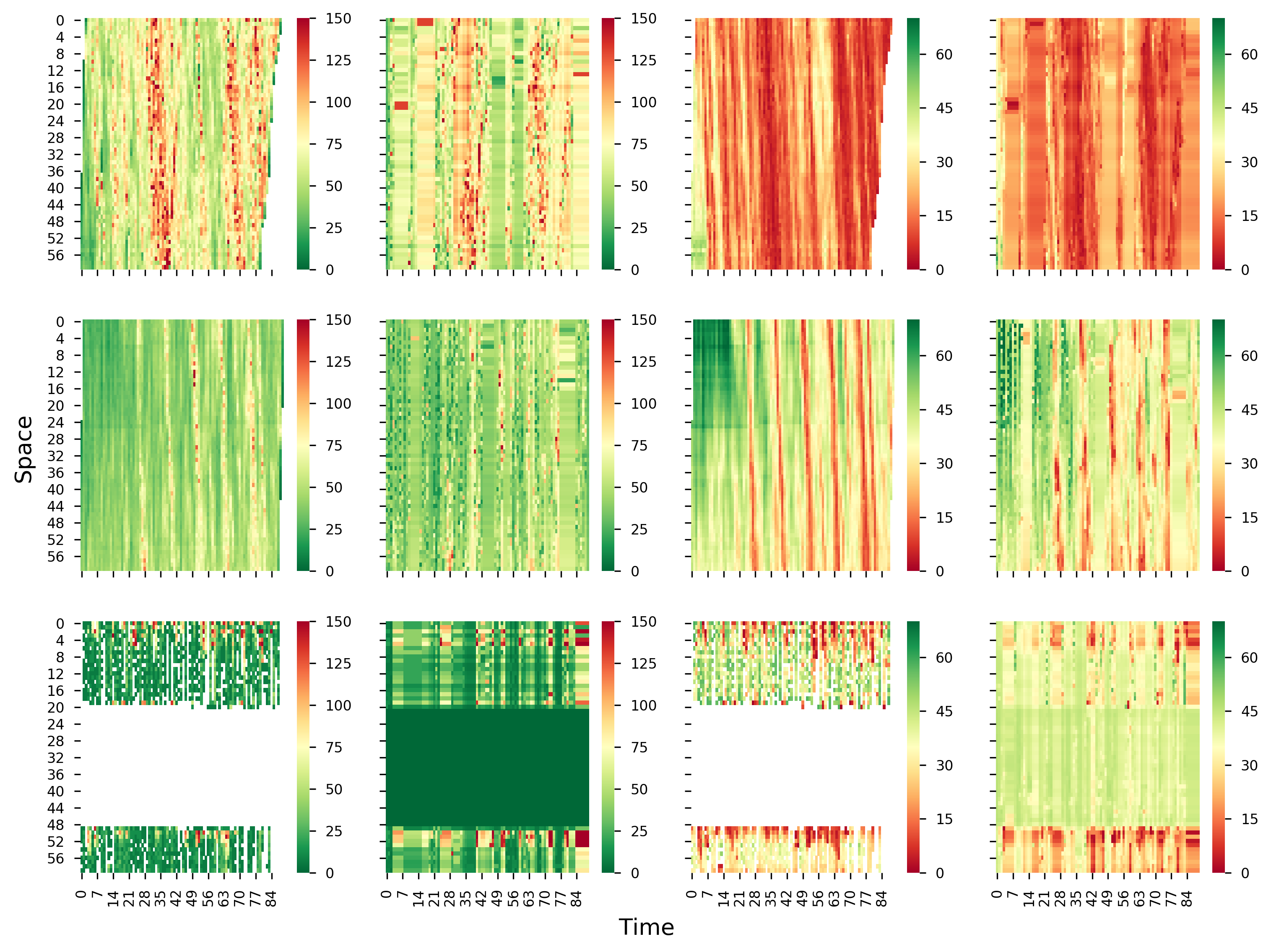}
	\caption{True and estimated density and speed for lane 4 \footnotesize{(first row: I-80, second row: US-101. third row: Lankershim Blvd; first column: ground true density, second column: estimated density, third column: ground true speed, fourth column: estimated speed; density unit: veh/km, speed unit: km/hour)}.}
	\label{fig:basic4}
\end{figure}

To visually inspect the estimation accuracy, we plot the true and estimated density and speed in time-space region for Lane 2 and Lane 4 in Figure~\ref{fig:basic2} and \ref{fig:basic4}. It can be seen that the estimated density and speed resemble the ground truth, even the congestion is discontinuous in the time-space region (see Lankershim Blvd in Figure~\ref{fig:basic2}). Again the Lane 4 of Lankershim Blvd is physically discontinuous, hence a large block of entries are entirely missing in time-space region (see the third row of Figure~\ref{fig:basic4}), and the blocked missingness may affect the proposed methods and increase the estimation errors.


\subsubsection{Regression Coefficients}
When estimating the speed, we use the LR2 for each lane separately. In this section, we look at the regression coefficients of the fitted Lasso model and interpret the coefficients from the perspective of the traffic flow theory. In particular, we select Lane 2 on US-101 and summarize the fitted coefficients in Table~\ref{tab:regression}. The regression coefficients for other lanes and networks  can be found in the supplementary materials.

\begin{table}[h]
	\centering
	\begin{tabular}{ccccccc}
		\toprule
		Variables & Coefficients  & Standard Error & t-statistic & p-value & $2.5\%$ quantile & $97.5\%$ quantile \\
		\midrule
		\texttt{Intercept}      &    0.0778  &    0.000  &  254.909   &   0.000   &    0.077   &    0.078\\
		\texttt{x1}         &   -0.4738  &    0.023  &  -20.676   &   0.000   &   -0.519   &   -0.429\\
		\texttt{x2}       &   -0.2630  &    0.018  &  -14.326   &   0.000   &   -0.299   &   -0.227\\
		\texttt{x3}         &   -0.4146  &    0.023  &  -17.700   &   0.000   &   -0.461   &   -0.369\\
		\texttt{x4}         &   -0.4781  &    0.024  &  -20.346   &   0.000   &   -0.524   &   -0.432\\
		\texttt{x5}         &   -0.2043  &    0.023  &   -9.030   &   0.000   &   -0.249   &   -0.160\\
		\texttt{x6}         &   -0.0947  &    0.017  &   -5.726   &   0.000   &   -0.127   &   -0.062\\
		\texttt{x7}         &   -0.1620  &    0.022  &   -7.352   &   0.000   &   -0.205   &   -0.119\\
		\texttt{x8}         &   -0.2354  &    0.022  &  -10.562   &   0.000   &   -0.279   &   -0.192\\
		\texttt{x9}         &   -0.1727  &    0.025  &   -6.790   &   0.000   &   -0.223   &   -0.123\\
		\texttt{x10}        &   -0.1788  &    0.019  &   -9.477   &   0.000   &   -0.216   &   -0.142\\
		\texttt{x11}        &   -0.1553  &    0.025  &   -6.198   &   0.000   &   -0.204   &   -0.106\\
		\texttt{x12}        &   -0.2067  &    0.025  &   -8.256   &   0.000   &   -0.256   &   -0.158\\
		\bottomrule
	\end{tabular}
	\vspace{0.5em}
	\caption{Coefficients of Lasso regression for Lane 2 on US-101 (\texttt{x1} to \texttt{x12} correspond to the number in Figure~\ref{fig:regression})}
	\label{tab:regression}
\end{table}

The R-square for Lane 2 on US-101 is 0.832, indicating the regression model is fairly accurate. From Table~\ref{tab:regression}, one can see the \text{intercept} is positive and it represents the free flow speed when the density is zero. Coefficients for \texttt{x1} to \texttt{x12} are all negative with high confidence, and this implies that higher density yields lower speed. 

Recall Figure~\ref{fig:regression}, suppose we want to estimate the density for cell 1, we refer to cell $1\sim4$ as the surrounding cells in the current lane and cell $5\sim12$ as the surrounding cells in the nearby lanes.
The coefficients of \texttt{x1} to \texttt{x4} are the most negative, indicating the densities of the surrounding cells in the current lane have the highest impact on the speed. The densities of surrounding cells in the nearby lanes also have negative impact on the speed but the magnitude is lower.

\subsection{Comparing different algorithms}
\begin{figure}[h]
	\centering
	\includegraphics[width=0.95\linewidth]{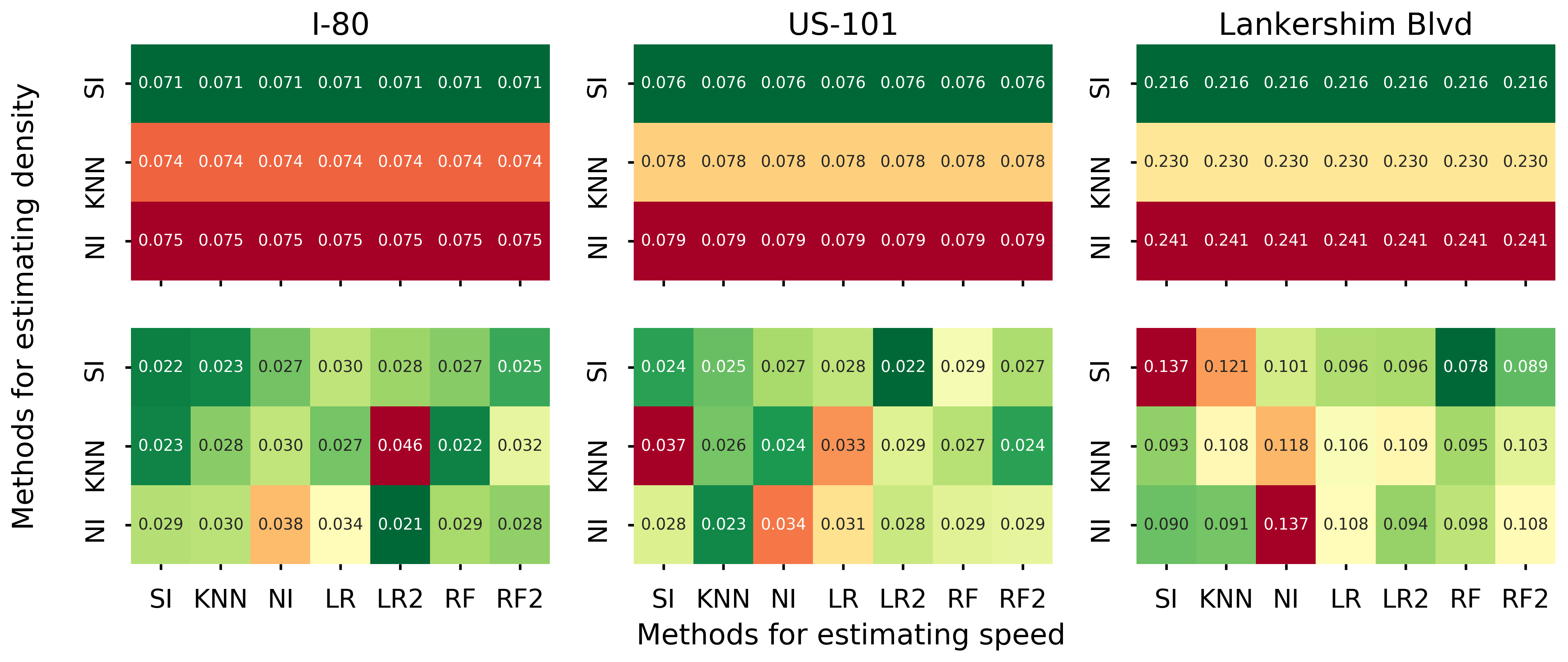}
	\caption{Average MSAPE1 for different estimation methods on each road.}
	\label{fig:alg}
\end{figure}

In this section, we examine different methods in estimating density and speed. Recall in section~\ref{sec:datadriven}, the matrix completion-based methods can estimate both density and speed while the regression-based methods can only estimate the speed. We run the proposed estimation method with different combinations of estimation methods for density and speed, and the rest of settings are the same as the baseline setting. To be precise, three methods are used to estimate density: Naive Imputation (NI), k-nearest neighbor imputation (KNN) and \textsc{SoftImpute} (SI). Seven methods will be used to estimate speed and they are NI, KNN, SI, LR1, LR2, RF and RF2. We plot the heatmap of MSAPE1 for each road separately, as presented in Figure~\ref{fig:alg}.

The speed estimation does not affect the density estimation as the density estimation is conducted first. SI always outperforms KNN and NI for density estimation. Different combination of algorithms perform differently on each road. We use A-B to denote the method that uses A for density estimation and B for speed estimation. NI-LR2 on I-80, SI-LR2 on US-101 and SI-RF on Lankershim Blvd outperform the rest of the methods in terms of speed estimation. Overall the SI-LR2 generates accurate estimation for all three roads.

\subsection{Impact of sensing power}
\begin{figure}[h]
	\centering
	\includegraphics[width=0.95\linewidth]{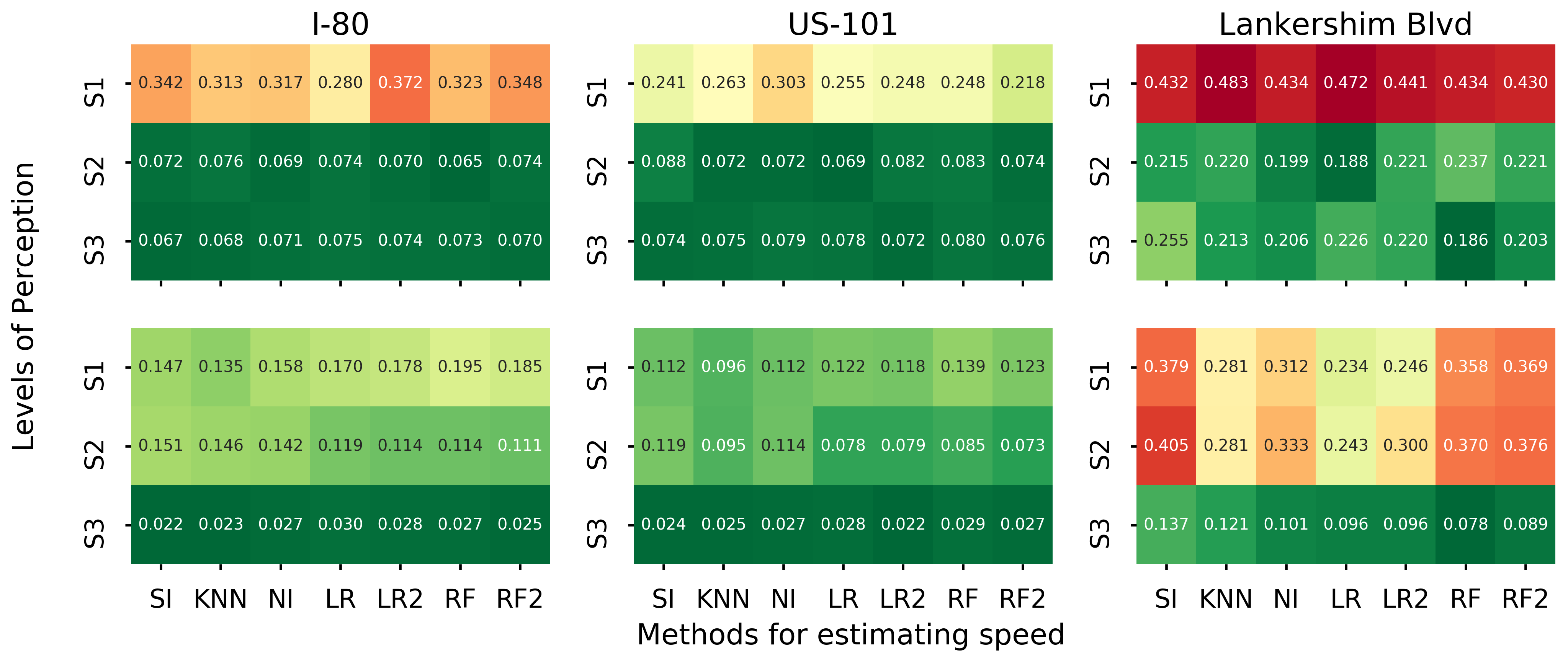}
	\caption{Estimation accuracy under three levels of perception \footnotesize{(first row: density, second row: speed)}.}
	\label{fig:levelp}
\end{figure}

We analyze the impact of sensing power of AVs on the estimation accuracy. Recall section~\ref{sec:lop} and \ref{sec:direct}, we consider three levels of perception for AVs. Based on Equation~\ref{eq:os1}, \ref{eq:os2} and \ref{eq:os3}, more entries in the time-space region get {\em directly} observed when the perception level increases. We run the proposed estimation method with different perception levels and different methods for speed estimation. Other settings are the same as the baseline setting. The heatmap of MSAPE1 for each road is presented in Figure~\ref{fig:levelp}.

As shown in Figure~\ref{fig:levelp}, the proposed methods performs the best on US-101 and the worst on Lankershim Blvd. With 5\% market penetration rate, at least S2 is required for I-80 and US-101 to obtain an accurate traffic state estimation. Similarly,  S3 is required for Lankershim Blvd to ensure the estimation quality. Later we will discuss the impact of market penetration rate on the estimation accuracy under different perception levels.

The estimation accuracy improves for all speed estimation algorithms and all three roads when the perception level increases. Different speed estimation algorithms perform differently on different roads within the same perception level. For example, in S2, the imputation-based methods outperform the regression-based method on I-80 and US-101, while the Lasso regression outperforms the rest on Lankershim Blvd. In S3, all the density estimation methods perform similarly on I-80 and US-101, while the regression-based method significantly outperform the imputation-based methods on Lankershim Blvd in terms of density estimation.

\subsection{Impact of AV market penetration rate}
\begin{figure}[h]
	\centering
	\includegraphics[width=0.95\linewidth]{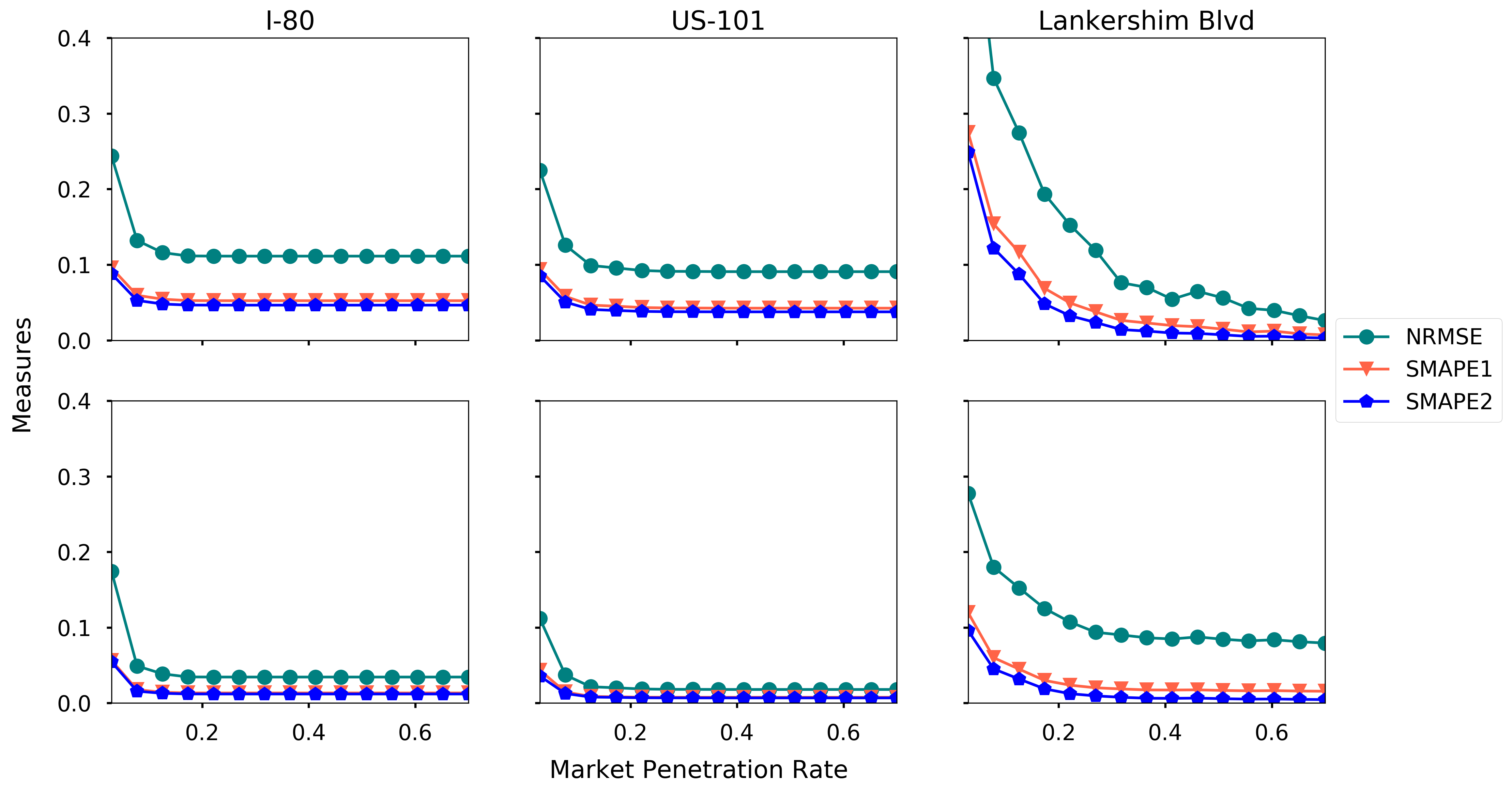}
	\caption{Estimation accuracy under different AV market penetration rate \footnotesize{(first row: density, second row: speed)}.}
	\label{fig:mp}
\end{figure}

To examine the impact of AV market penetration rate, we run the proposed method with different market penetration rates ranging from $0.03$ to $0.7$, and the rest of settings are the baseline setting. The experiment results are presented in Figure~\ref{fig:mp}.

\begin{figure}[h]
	\centering
	\includegraphics[width=0.95\linewidth]{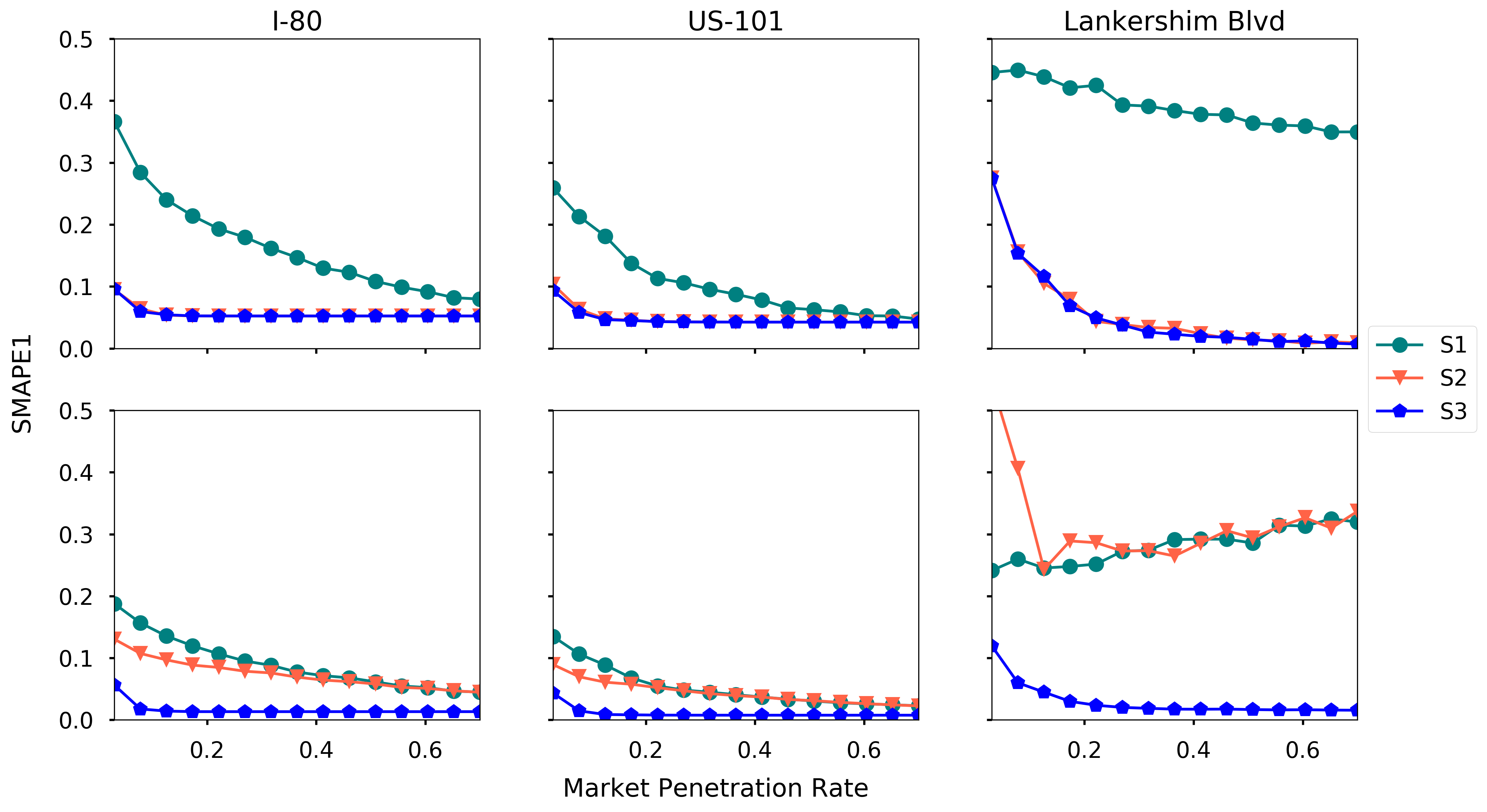}
	\caption{SMAPE1 under different market penetration rates and perception levels \footnotesize{(first row: density, second row: speed)}.}
	\label{fig:mp2}
\end{figure}

Generally the estimation accuracy increases when the AV market penetration rate increases. $5\%$ penetration rate is large enough for an accurate estimation for I-80 and US-101, while Lankershim Blvd requires larger penetration rate. To further investigate the impact of market penetration rate under different levels of perception, we run the experiment with different penetration rate under three levels of perception, and the results are presented in Figure~\ref{fig:mp2}.

One can read that S2 and S3 yield the same density estimation as the vehicle detection is enough for density estimation. Better speed estimation can be achieved on S3 since more vehicles are tracked and the speeds are measured. Again, Figure~\ref{fig:mp2} indicates at least S2 is required for I-80 and US-101 to obtain an accurate traffic state estimation, and S3 is required for Lankershim Blvd to ensure the estimation quality.

In addition to above findings, another interesting finding is that when the market penetration rate is low, the regression methods usually outperform the matrix completion-based methods, while the matrix completion-based methods outperform the regression-based methods when the market penetration rate is high.
\subsection{Platooning}
In the baseline setting, AVs are uniformly distributed in the fleet, while many studies suggest that a dedicated lane for platooning can further enhance mobility \citep{ramezani2017capacity}. In this case, AVs are not uniformly distributed on the road. To simulate the dedicated lane, we view all vehicles on Lane 1 of I-80, Lane 1 of US-101, and Lane 1,2 of Lankershim Blvd are AVs, and all the vehicles on other lanes are conventional vehicles. To compare, we also set another scenario with the same number of vehicles, which are treated as AVs, uniformly distributed on roads. We run the proposed method on both scenarios with the rest of settings being the baseline setting, and the results are presented in Table~\ref{tab:plat}.

\begin{table}[h]
	\centering
	\begin{tabular}{lllllll}
		\toprule
		\multirow{2}{*}{Measures} & \multicolumn{3}{c}{Density}  & \multicolumn{3}{c}{Speed}\\
		\cmidrule(r){2-7}
		~    & NRMSE & SMAPE1 &  SMAPE2 & NRMSE & SMAPE1 &  SMAPE2 \\
		\midrule
		I-80 & 31.71/31.65 & 12.45/12.47 & 11.86/11.85 & 19.78/19.84 & 8.08/78.12 & 6.92/6.95\\
		US-101 & 25.93/25.91 & 10.18/10.16 & 9.62/9.61 & 15.13/15.15 & 6.05/6.05& 4.90/4.91\\
		Lankershim & 66.67/66.71 & 31.17/31.58 & 28.89/28.70 & 38.99/37.60 & 20.03/19.74& 16.00/15.67\\
		\bottomrule
	\end{tabular}
	\vspace{0.5em}
	\caption{Estimation accuracy with AVs on dedicated lanes and uniformly distribution \footnotesize{(A/B: A is for the uniformly distribution scenario, B is for the dedicated lane scenario).}}
	\label{tab:plat}
\end{table}

As can be seen from Table~\ref{tab:plat}, the distribution of AVs has marginal impact on the estimation accuracy. The proposed method performs similarly on the scenarios of the dedicated lane and uniformly distribution for all three roads, which is probably because the detection range of LiDAR is large enough to cover the width of the roads.

\subsection{Sensitivity analysis}
In this section, we examine the sensitivity of various factors ({\em e.g.} LiDAR detection range, sampling rate, detection missing rate, and speed detection noise) in our experiments. 

\begin{figure}[h]
	\centering
	\includegraphics[width=0.95\linewidth]{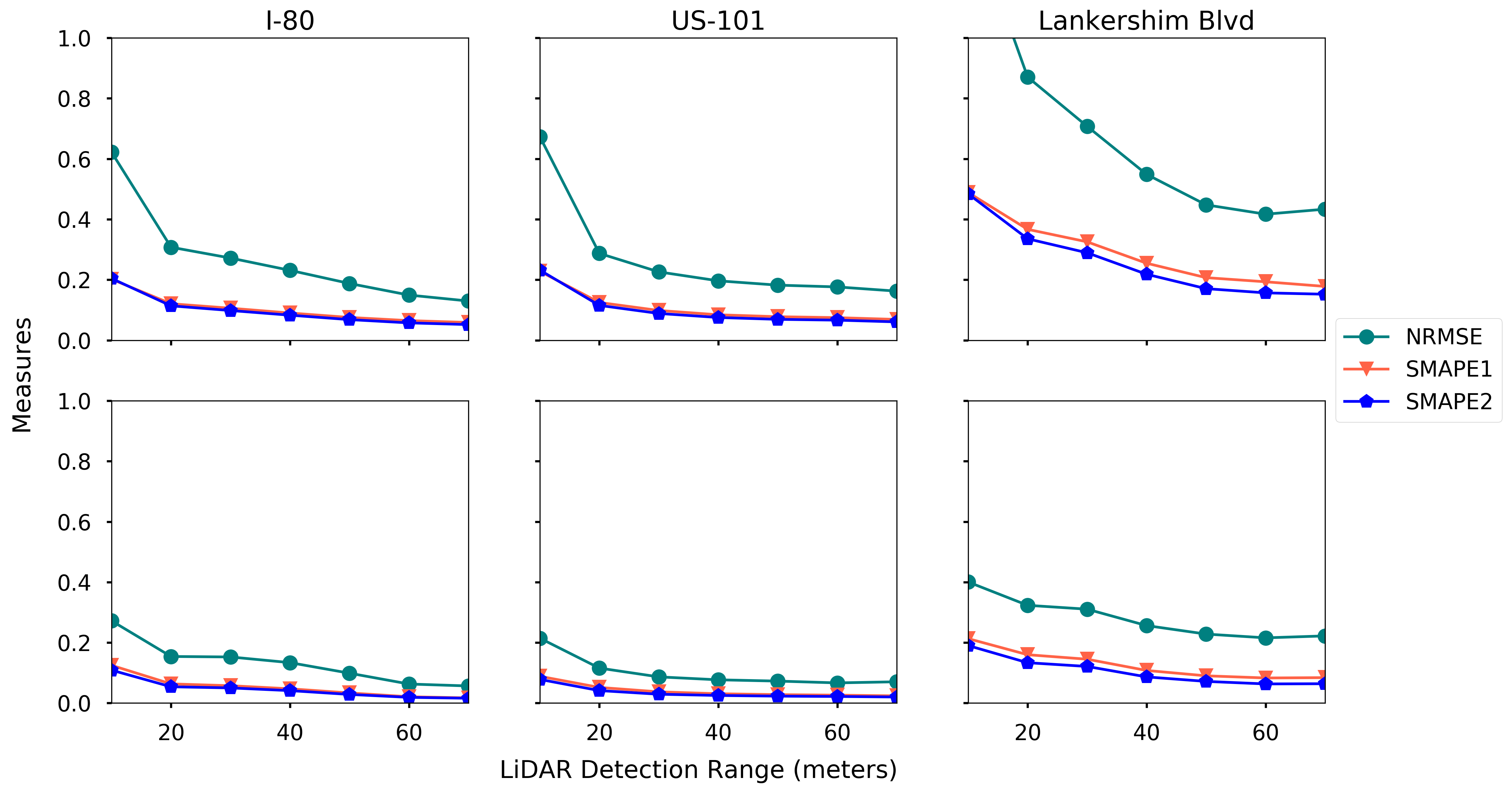}
	\caption{Estimation accuracy with different LiDAR detection range \footnotesize{(first row: density, second row: speed)}.}
	\label{fig:meter}
\end{figure}

{\bf LiDAR detection range.} The detection range of LiDAR varies in a wide range for different brands \citep{lidar}. We run the proposed estimation method using different detection rage ranging from 10 meters to 70 meters, and the rest of settings are the baseline setting. The estimation accuracy for each road is presented in Figure~\ref{fig:meter}.

One can read that the estimation error reduces for both density and speed when the detection range increases. The gain in estimation accuracy becomes marginal when the detection range is large. For example, when the detection range exceed 40 meters on US-101, the improvement of the estimation accuracy is negligible. Another interesting observation is that, on Lankershim Blvd, even 70-meter detection range cannot yield a good density estimation with 5\% market penetration rate.

\begin{figure}[h]
	\centering
	\includegraphics[width=0.95\linewidth]{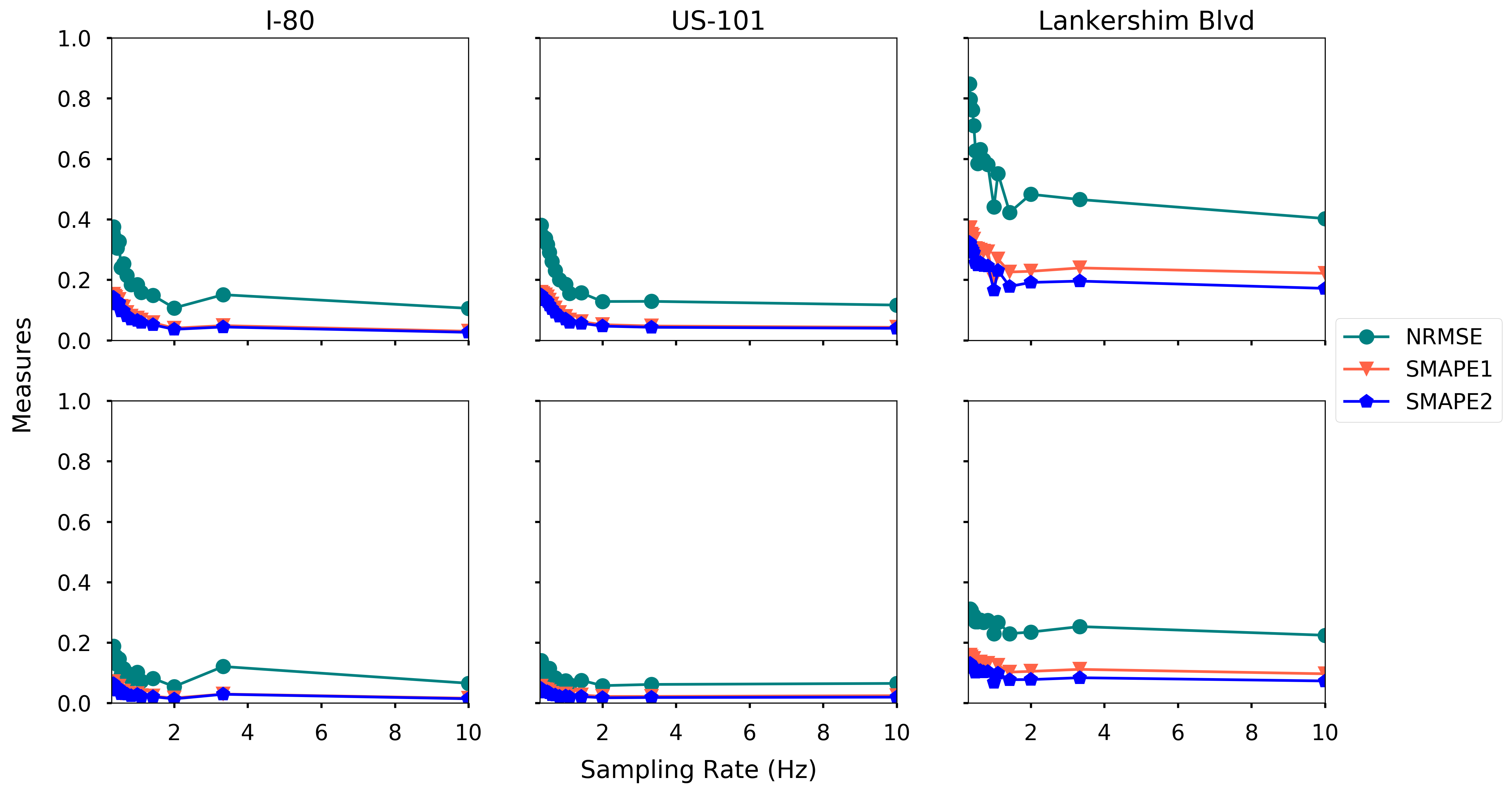}
	\caption{Estimation accuracy with different sampling rate \footnotesize{(first row: density, second row: speed)}.}
	\label{fig:sr}
\end{figure}

{\bf Sampling rate.} Recall that the sampling rate denotes the frequency of the message (which contains the location/speed of itself and detected vehicles) to the data center, as discussed in section~\ref{sec:datacenter}. When the sampling rate is low, we conjecture that the data center received fewer messages, which increases the estimation error. To verify our conjecture, we run the proposed estimation method with different sampling rate ranging from 0.3Hz to 10Hz, and the rest of settings are the base setting. The estimation accuracy on each road is further plotted in Figure~\ref{fig:sr}.

The estimation accuracy increases when the sampling rate increases for all three roads, as expected. The density estimation is more sensitive to the sampling rate than the speed estimation. This is probably because the density changes dramatically in time-space region, while the speed is relatively stable. 

\begin{figure}[h]
	\centering
	\includegraphics[width=0.95\linewidth]{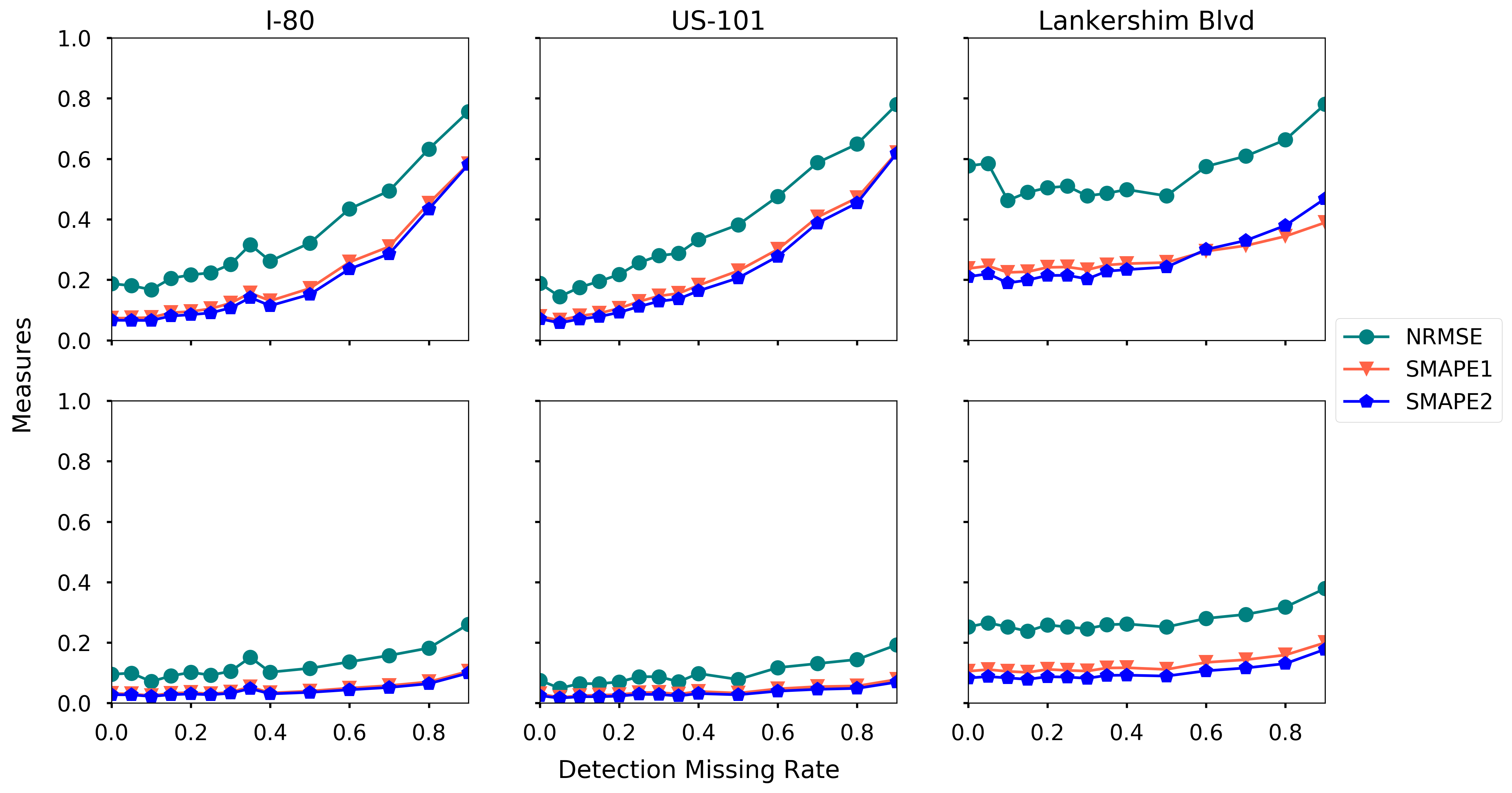}
	\caption{Estimation accuracy under different detection missing rate \footnotesize{(first row: density, second row: speed)}.}
	\label{fig:missing}
\end{figure}

{\bf Detection missing rate.} The AVs might overlook a certain vehicle during the detection, and we use the missing rate to denote the probability. We examine the impact of the missing rate by running the proposed estimation method with different missing rate ranging from  0.01 to 0.9, and the rest of settings are the baseline setting. We plot the estimation accuracy for each road separately, as presented in Figure~\ref{fig:missing}.

From Figure~\ref{fig:missing} one can read that the estimation error increases when the missing rate increases for all three roads.  The density estimation is much more sensitive to the missing rate than the speed estimation. This is because that overlooking vehicles has a significant impact on density estimation, while speed estimation only needs a small fraction of vehicles being observed. 

\begin{figure}[h]
	\centering
	\includegraphics[width=0.95\linewidth]{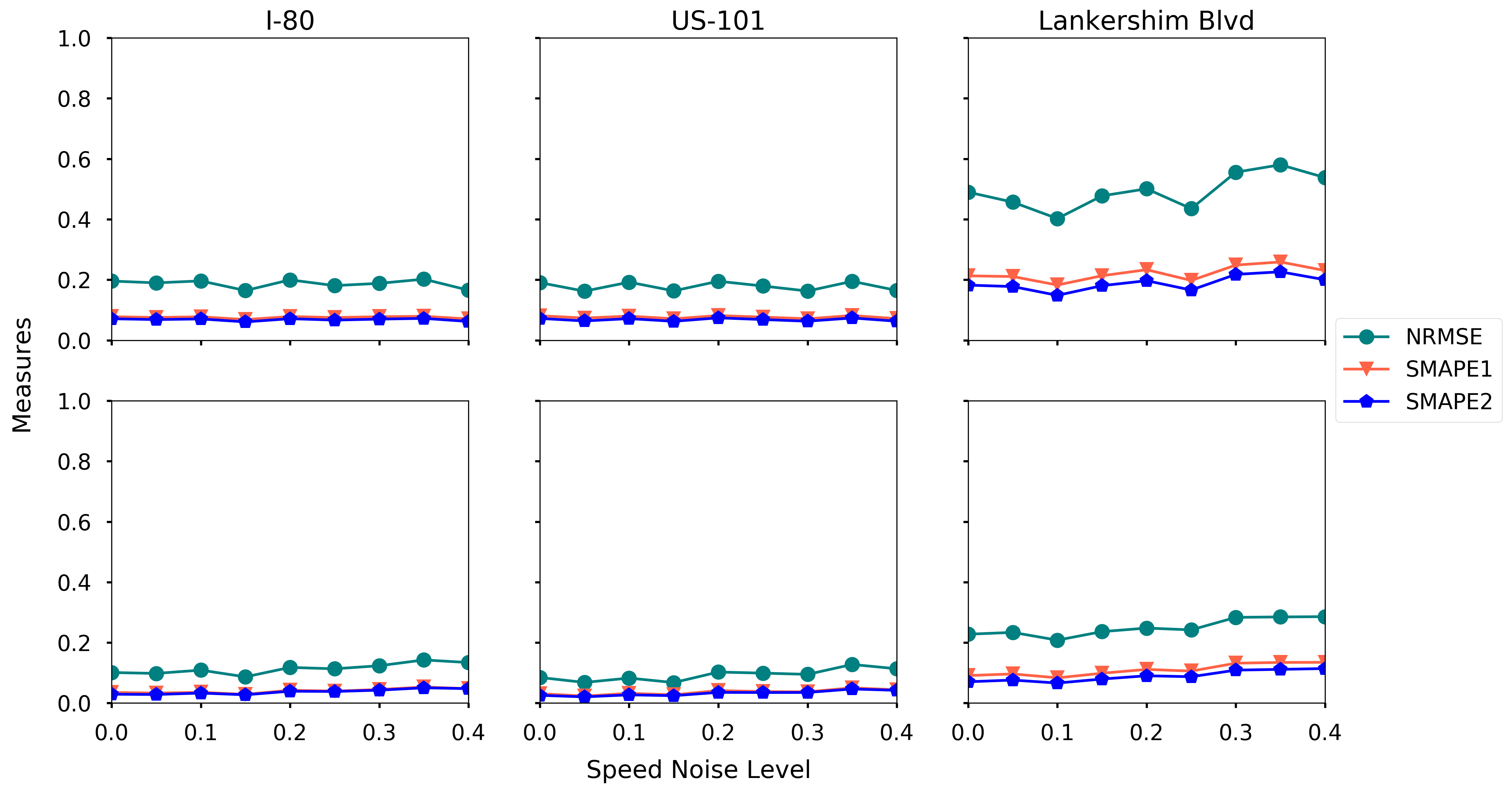}
	\caption{Estimation accuracy under different levels of speed noise \footnotesize{(first row: density, second row: speed)}.}
	\label{fig:spdnoise}
\end{figure}

{\bf Noise level in speed detection.} We further look at the impact of noise in speed detection. We assume that the speed of a certain vehicle is detected with noise, and the noise level is denoted as $\xi$. If the true vehicle speed is $\nu$, we sample $\bar{\xi}$ from the uniform distribution $\texttt{Unif}(-\xi, \xi)$, and then the detected vehicle speed is assumed to be $\nu + \nu\bar{\xi}$. We run the proposed estimation method by sweeping $\xi$ from $0.0$ to $0.4$, and the rest of settings are baseline setting. The estimation accuracy is presented in Figure~\ref{fig:spdnoise}.

Surprisingly the proposed method is robust to the noise in speed detection, as the estimation errors remain stable when the speed noise level increases. One explanation for this is that the speed of each cell is computed by averaging the detected speeds from multiple vehicles, hence the detection noise is complemented and reduced based on the Law of large numbers.

\clearpage
\section{Conclusions}
\label{sec:con}

This paper proposes a high-resolution traffic sensing framework with autonomous vehicles (AVs). The framework leverages the perception power of AVs to estimate the fundamental traffic state variables, namely, flow, density and speed, and the underlying idea is to use AVs as moving observers to detect and track vehicles surrounded by AVs. We discuss the potential usage of each sensor mounted on AVs, and categorize the sensing power of AVs into three levels of perception. Then the data-driven traffic sensing framework is rigorously formulated. The proposed framework consists of two steps: 1) directly observation of the traffic states using AVs; 2) data-driven estimation of the unobserved traffic states. In the first step, we define the direct observations under different perception levels. The second step is done by estimating the unobserved density using matrix-completion methods, followed by the estimation of unobserved speed using either matrix-completion methods or regression-based methods. The implementation details of the whole framework are further discussed. 

The Next Generation Simulation (NGSIM) data is adopted to examine the accuracy and robustness of the proposed framework. The proposed estimation framework is examined extensively on I-80, US-101 and Lankershim Boulevard. In general, the proposed framework estimates the traffic states accurately with a low AV market penetration rate. The speed estimation is always easier than density estimation, as expected. Results show that, with 5\% AV market penetration rate, at least S2 is required for I-80 and US-101 to obtain an accurate traffic state estimation, while S3 is required for Lankershim Blvd to ensure the estimation quality. During the estimation of speed, all the coefficients in the Lasso regression can be interpreted by the traditional flow theory. In addition, sensitivity analysis regarding AV penetration rate, sensor configuration, speed detection noise, perception accuracy was conducted.

This study will help policymakers and private sectors (e.g Uber, Waymo) understand the values of AVs in traffic operation and management, especially the values of massive data collected by AVs. Hopefully, new business models to commercializing the data \citep{mobilityeco} or collaborations between private sectors and public agencies can be triggered. In the near future, we will examine the sensing capabilities of AVs at network level and extend the proposed traffic sensing framework to large-scale networks. Another interesting research direction is to investigate the privacy issue when AVs share the observed information with the data center.

\section*{Supplementary Materials}
The proposed traffic sensing framework is implemented in Python and open-sourced on Github\footnote{\url{https://github.com/Lemma1/NGSIM-interface}}. Readers can reproduce all the experiments in section~\ref{sec:exp}. Additionally, the Github repository also contains some analysis that is omitted in section~\ref{sec:exp}.

\section*{Acknowledgments}
This research is funded in part by Traffic 21 Institute and Carnegie Mellon University's Mobility 21, a National University Transportation Center for Mobility sponsored by the US Department of Transportation. 

\cleardoublepage
\bibliography{report}
\cleardoublepage

\end{document}